\documentclass[twocolumn]{aastex701}
\usepackage{amsmath}

\usepackage{xspace}
\usepackage{subcaption}

\newcommand{\thisgrborig}{GRB\,160325A\xspace}
\newcommand{\thisgrb}{GRB\,160325A\ensuremath{^\prime}\xspace}

\providecommand{\sw}[1]{\texttt{#1}}

\newcommand{\threeml}{3ML\xspace}
\newcommand{\polarpy}{\sw{polarpy}\xspace}
\newcommand{\polpy}{PolPy\xspace}

\newcommand{\fermi}{\emph{Fermi}\xspace}
\newcommand{\integral}{\emph{INTEGRAL}\xspace}
\newcommand{\polar}{\emph{POLAR}\xspace}
\newcommand{\polarb}{\emph{POLAR-2}\xspace}
\newcommand{\cosi}{\emph{COSI}\xspace}
\newcommand{\astrosat}{\emph{AstroSat}\xspace}
\newcommand{\daksha}{\emph{Daksha}\xspace}

\begin{document}
\title{\polpy: A universal tool for X-ray/gamma-ray polarimetry using common data format standards }

\correspondingauthor{Sujay Mate}
\email{sujay.mate@gmail.com, sujay.mate@nasa.gov}

\correspondingauthor{Hancheng Li}
\email{Hancheng.Li@unige.ch}

\author[orcid=0000-0001-5536-4635]{Sujay Mate}
\affiliation{Raman Research Institute, C. V. Raman Avenue, Sadashivanagar, Bangalore, Karnataka, 560080, India}
\affiliation{NASA Postdoctoral Program Fellow, Astrophysics Science Division, NASA Goddard Space Flight Center, 8801 Greenbelt Road, Greenbelt, MD 20771, USA}
\email{sujay.mate@nasa.gov}
\email{sujay.mate@gmail.com}

\author[0000-0002-5963-1494]{Hancheng Li}
\affiliation{Department of Astronomy, University of Geneva, 16 Chemin
d’Ecogia, Versoix, CH-1290, Switzerland}
\email{Hancheng.Li@unige.ch}

\author[0009-0002-7897-6110]{Utkarsh Pathak}
\affiliation{Department of Physics, Indian Institute of Technology Bombay, Powai, Mumbai, Maharashtra 400076, India}
\email{utkarshpathak.07@iitb.ac.in}

\author[0009-0005-5080-0107]{Yashowardhan Rai}
\affiliation{Department of Physics, Indian Institute of Technology Bombay, Powai, Mumbai, Maharashtra 400076, India}
\email{yash_rai@iitb.ac.in}

\author[0000-0002-2498-0213]{Nicolas De Angelis}
\affiliation{INAF Istituto di Astrofisica e Planetologia Spaziali, Via del Fosso del Cavaliere 100, 00133 Roma, Italy}
\email{nicolas.deangelis@inaf.it}

\author[0000-0002-6112-7609]{Varun Bhalerao}
\affiliation{Department of Physics, Indian Institute of Technology Bombay, Powai, Mumbai, Maharashtra 400076, India}
\email{varunb@iitb.ac.in}

\author[0000-0003-0441-4959]{Merlin Kole}
\affiliation{Space Science Center, University of New Hampshire, Durham, NH 03824, USA}
\email{merlinkole@gmail.com}

\begin{abstract}
X-ray/gamma-ray polarimetry has been a rapidly developing field in recent years. However, the lack of standardized analysis frameworks has hindered cross-instrument comparisons and joint multi-mission studies. To address this, we present \polpy, a universal tool designed for polarimetry of high-energy astrophysical transients. Built as a plugin for the Multi-Mission Maximum Likelihood (3ML) framework, \polpy introduces standardized data and response formats, common reference frames based on IAU conventions for defining polarization angles, and a robust likelihood-based template-matching algorithm. The article briefly describes these standards and presents tests and results performed to validate the software using simulated gamma-ray burst (GRB) injections across two different mission mass models: \polar and \daksha. Our verification tests demonstrate that \polpy accurately recovers the injected polarization angles and fractions without systematic bias. Furthermore, we show that performing simultaneous joint fits across multiple detectors and instruments can significantly improve parameter constraints compared to single-instrument analyses. By unifying data formats and methodology, \polpy enables reliable joint polarimetric analysis for current and future high-energy polarimeters.
\end{abstract}



 \section{Introduction}
In recent years, X-ray/gamma-ray polarimetry has gained renewed interest with the advent of sensitive detector technologies capable of overcoming the challenges posed by the ``photon-hungry'' nature of polarimetry. The Imaging X-ray Polarimetry Explorer~\citep[IXPE,][]{Weisskopf2022} has made strides in advancing our knowledge of high-energy persistent sources such as pulsars, magnetars, X-ray binaries and active galactic nuclei through polarimetric measurements in the soft X-ray band. In the hard X-ray and soft gamma-ray bands, missions like \polar~\citep{Produit2018,Kole2020} and the Cadmium Zinc Telluride Imager (CZTI) onboard \astrosat~\citep{Bhalerao2017a,Chattopadhyay2022} have provided insights into the prompt emission of Gamma-Ray Bursts (GRBs), while \integral\ observatory has also contributed to polarimetric measurements of GRBs and other high-energy sources~\citep[e.g.,][]{Dean2008, Gotz2009, Laurent2011}. This field is expected to develop further with the impending launch of the Compton Spectrometer and Imager~\citep[\cosi;][]{Tomsick2024}, \polarb~\citep{Kole2025} and proposed missions like \daksha~\citep{Bhalerao2024,Bhalerao2024a,Bala2023}. These missions will provide a wealth of polarimetric data in the next decade, which will be crucial for understanding the physics behind GRB prompt emission or transient events like magnetar bursts/flares \citep{Toma2009}.

As the field of high-energy polarimetry is still in its nascent stage, the data analysis methods and frameworks have not matured as they have for spectroscopic or timing analysis. Furthermore, the physics and detectors for measuring polarization vary for different energy bands --- for instance, photo-electric polarimeters are needed in soft X-rays while scattering polarimeters are needed in hard X-ray/gamma-ray regimes --- each needing a different analysis method. Polarization measurements are still limited by low statistics in most cases, especially for transient sources. These challenges have led to the development of instrument-specific analysis software and methods. The downside of this is that it becomes difficult to compare results across instruments and to perform joint analysis of data from multiple instruments. Both these limitations became apparent in the case of GRB polarimetry when \polar and \astrosat/CZTI results were published~\citep{Kole2020, Chattopadhyay2022}. At first glance, the results seem to be inconsistent: \polar results reported low or unconstrained polarization fraction for all 14 GRBs they analyzed, while in the 20-GRB sample of \astrosat/CZTI, a few were found to have higher polarization. The instruments were contemporaries, and two GRBs were co-detected by the pair --- but differences in the data analysis pipelines made it impossible to perform a joint fit that would have validated the techniques and helped understand the differences in the results.

To overcome these challenges and to set standards for any future polarimetric analysis across missions, we have developed \polpy, a plugin for the Multi-Mission Maximum Likelihood framework~\citep[\threeml;][]{Vianello2016}, that provides a unified framework for the analysis of polarimetric data. This article describes the technical details of the \polpy framework, including the current data and response standards, common reference frame definitions and the implementation of the likelihood-based fitting approach. We also demonstrate the ability of the tool to perform joint polarimetric analysis of GRBs using simulated data, and applications to real data will be published in a future manuscript (Pathak et al in prep). The article is structured as follows: In Section~\ref{sec:overview}, we present the architecture of the framework, including the data, response and common reference frame definitions. In Section~\ref{sec:validation}, we present the results of our validation tests carried out using GRB injections across two different missions. In Section~\ref{sec:discussion}, we discuss future use of \polpy and potential improvements in future versions, followed by a summary in Section~\ref{sec:conclusion}.

\section{\polpy: Overview}~\label{sec:overview}
The field of hard X-ray polarization measurements is currently dominated by \polar and \astrosat/CZTI; hence, it was natural that these two became the starting point of our development. \polar data are analyzed by the \polarpy tool developed by the mission team~\citep[for more details, see][]{Burgess2019,Kole2020}. \polarpy relies on pre-computing Azimuthal Scattering Angle Distribution (ASAD) ``templates'' in the Polarization Angle (PA hereafter) and Polarization Fraction (PF) space, and forward folding them to match the observed ASAD. Data from \astrosat/CZTI are currently analyzed using the modulation curve fitting method \citep{Chattopadhyay2014,2015A&A...578A..73V}, but the team has also demonstrated the template-matching method for their analysis~\citep{aarthy2021multiwavelength}. A similar template matching approach has also been adopted by the proposed \daksha mission~\citep{Bala2023}. Hence, we selected the template-based approach as the common algorithm for the polarimetric analysis using \polpy. This choice also naturally simplified many development aspects as we were able to reuse some of the code base developed for \polarpy. In this section, we describe the important details about the \polpy framework. The detailed documentation can be found by installing \polpy via pip and/or by visiting the GitHub repository\footnote{\url{https://pypi.org/project/polpy/}\\\url{https://github.com/threeML/polpy/}}.

\subsection{Common Reference Frames}~\label{sec:ref_frames}
A key challenge in doing cross-instrument polarization analysis, compared to cross-instrument spectral or timing analysis, is to have a universal reference frame to define the polarization angle and correctly transform the templates that are computed in a local reference frame of the instrument. In order for this transform to work across instruments, it is also important to define a standard definition of the ``local" frame in which the templates are computed. Here we give details about these frames, and all the necessary transformations that are needed to correctly transform templates to a uniform frame in order to fit the data. Details about how to compute the individual transformation matrices are given in the \polpy documentation.
\begin{figure*}[!htbp]
    \centering
    \includegraphics[width=\textwidth]{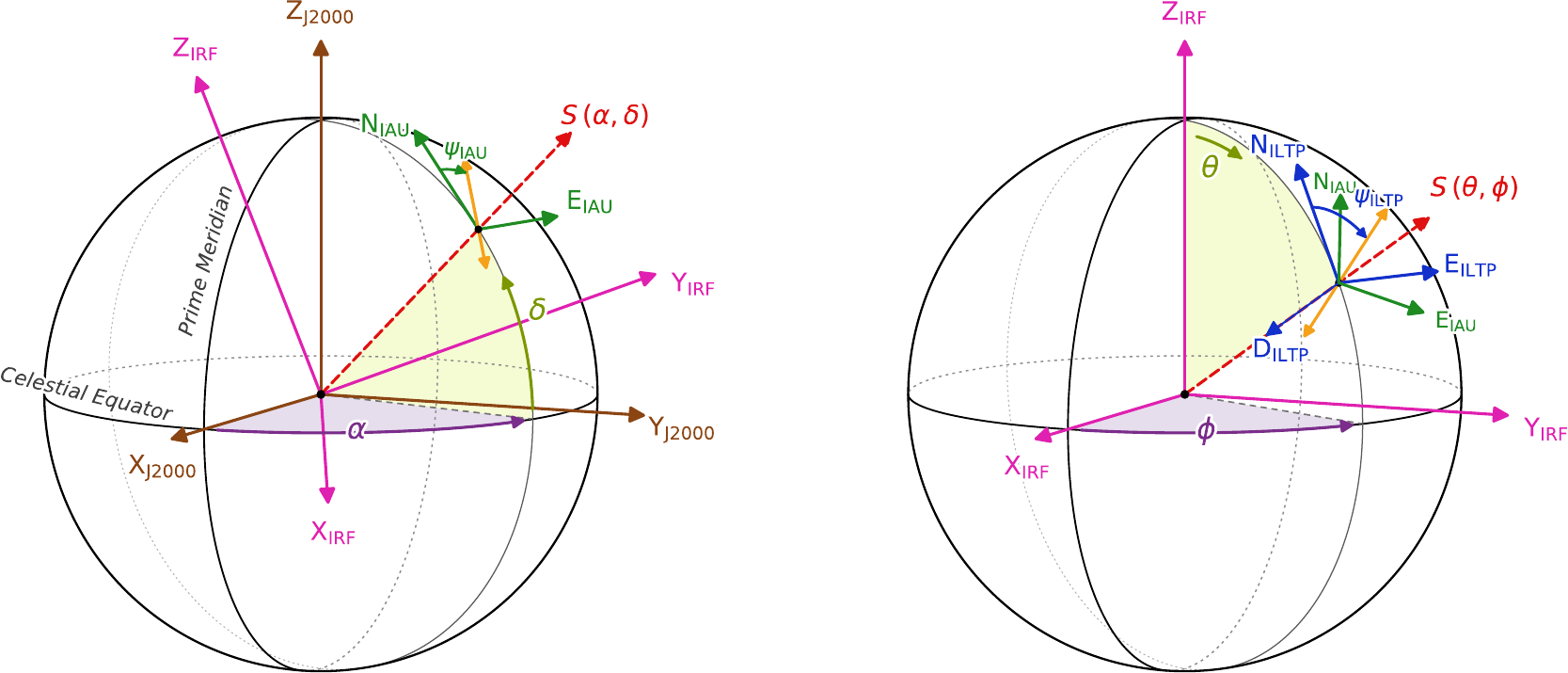}
\caption{Schematic definitions of the reference frames used in \polpy. 
\textbf{Left:} Using the celestial coordinate frame ($X_{\text{J2000}}, Y_{\text{J2000}}, Z_{\text{J2000}}$) as the base reference, the figure shows the relation between the IAU frame ($N_{\text{IAU}}, E_{\text{IAU}}$), and the Instrument Reference Frame (IRF; $X_{\text{IRF}}, Y_{\text{IRF}}, Z_{\text{IRF}}$) using the local tangent plane. Following the IAU convention, at source position $S(\alpha, \delta)$, the polarization angle $\psi_{\text{IAU}}$ is measured positively from the local North ($N_{\text{IAU}}$) towards the local East ($E_{\text{IAU}}$). \textbf{Right:} The figure shows the relation between the IRF and the Instrument Local Tangent Plane (ILTP; $N_{\text{ILTP}}, E_{\text{ILTP}}, D_{\text{ILTP}}$) frame in which the polarization responses are computed. The ILTP follows similar convention as IAU to define the polarization angle ($\psi_{\text{ILTP}}$) at the source position $(\theta, \phi)$ relative to the IRF. Note that both $\psi_{\text{IAU}}$ and $\psi_{\text{ILTP}}$ are essentially in the same plane and have an offset between them (see Figure~\ref{fig:ref_frame_transform}).}
\label{fig:ref_frame_def}
\end{figure*}
\begin{figure}[b]
    \centering
    \includegraphics[width=0.7\linewidth]{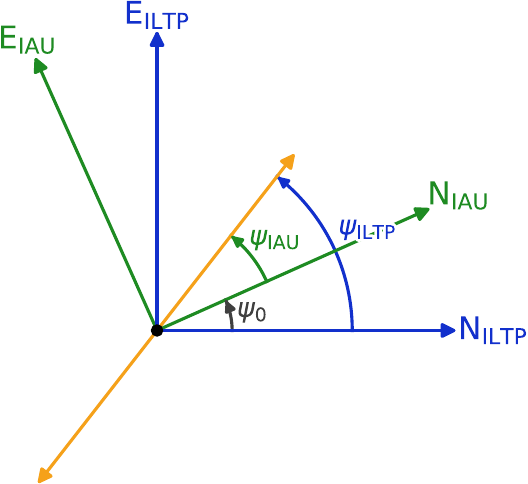}
\caption{Relation between the ILTP and IAU frames projected onto the plane perpendicular to the source direction with respective polarization angles ($\psi_\mathrm{ILTP}$ and $\psi_\mathrm{IAU})$ and the offset angle ($\psi_0$) marked appropriately.}\label{fig:ref_frame_transform}
\end{figure}

Figure~\ref{fig:ref_frame_def} shows all the reference frames that are in use in \polpy. There are total of four coordinate frames which are used by \polpy. Two of them are global, same across any instrument; while other two are specific to an instrument (local), but have a fixed transformation between them. The universal reference frame for the fitting and polarization angle is chosen to be the one defined by the 1973 IAU resolution\footnote{\url{https://www.iau.org/IAU/Iau/Publications/List-of-Resolutions.aspx}~(year 1973, commission 40, page 21).}, which is also the standard definition used across optical and radio astronomy. It defines the zero of the polarization angle to be at the ``local'' North with respect to the celestial reference frame (J2000 hereafter) and it increases positively towards the ``local'' East (i.e. counter-clockwise when the source is viewed from a point on the Earth). We denote this reference frame by the suffix `IAU' hereafter, and the PA is denoted by $\psi_\mathrm{IAU}$. The second key reference frame is the Instrument Reference Frame (IRF). The right panel of Figure~\ref{fig:ref_frame_def} shows the source located on the celestial sphere, but now with axes aligned to the IRF. Similar to the IAU convention, we now use the local tangent plane to define a coordinate frame, called Instrument Local Tangent Plane (ILTP) frame, with the axes $N_\text{ILTP}$ and $E_\text{ILTP}$, and the PA as $\psi_\mathrm{ILTP}$.

It should be noted that both the IAU and ILTP frames are in the same plane (projected sky plane when viewed from the Earth) and they differ only by an offset ($\psi_0$) defined by the source position and the pointing direction of the instrument (z-axis of the IRF) in the celestial frame. Figure~\ref{fig:ref_frame_transform} shows this projection and the relation between the two frames\footnote{Note that \polar’s source coordinate system (the local incidence frame) was originally defined in a similar manner in \citep{2022MNRAS.512.2827L}. However, a slight difference in convention makes the local polarisation angle $\psi_\mathrm{S}$ supplementary to $\psi_\mathrm{ILTP}$, such that $\psi_\mathrm{S} + \psi_\mathrm{ILTP} = 180^\circ$. For \polarb~\citep{Kole2025}, the revised definition of $\psi_\mathrm{ILTP}$ will be adopted in the data analysis.}. Mathematically, they are related as:
\begin{equation}
    \psi_{\text{IAU}} = \psi_{\text{ILTP}} - \psi_0,
\end{equation}

Assuming these definitions, \polpy internally converts the polarization response matrix from ILTP to IAU by computing $\psi_0$ for each instrument, and performs the fit in the IAU frame. The PA posterior produced by \polpy has the same reference as the IAU convention, and the results can be quoted directly without a need to apply an instrument-specific offset. However, this implies that to obtain correct results, it is critical that the instrument response (see Section~\ref{sec:rsp_format}) is created by strictly following the above definitions.

\subsection{Data and Response Format}~\label{sec:formats}
To enable cross-instrument polarimetric modeling with \polpy, we standardize both data and response formats following HEASARC OGIP conventions \citep[CAL/GEN/92-002;][]{george1992calibration}. Because GRB polarimeters employ diverse event-selection criteria to identify coincident scattering pairs, we did not include this process within the scope of \polpy. Instead, we adopt a reduced, instrument-agnostic event structure together with a multi-extension response format, both of which are discussed below. Here, we only give important details about the data and response formats. More details that would be needed to create these files are discussed in the \polpy documentation\footnote{\url{https://github.com/threeML/polpy/tree/master/docs}}.

\subsubsection{Polarization Data Format (\texttt{.pevt})}~\label{sec:data_format}
Event-level data are stored in a single FITS binary table extension called \texttt{POLEVENTS}. Each row in this extension represents a validated event pair defined by four mandatory columns: an event timestamp (\texttt{TIME}), a raw pulse-height analyzer channel (\texttt{CHANNEL}), a scattering-angle bin index at maximum detector resolution (\texttt{SABIN}), and the fractional dead time associated with the event (\texttt{DEADFRAC}). The extension header carries standard OGIP keywords (\texttt{TELESCOP}, \texttt{INSTRUME}, \texttt{EMIN}, \texttt{EMAX}, \texttt{NCHANS}, \texttt{NSABINS}), together with the instrument attitude and target position keywords (\texttt{RAX}, \texttt{DECX}, \texttt{RAZ}, \texttt{DECZ}, \texttt{RAGRB}, \texttt{DECGRB}, all in degrees) required by \polpy to correctly transform the responses created in the ILTP frame to the IAU frame as discussed in Section~\ref{sec:ref_frames}.

\subsubsection{Polarization Response Format (\texttt{.prsp})}~\label{sec:rsp_format}
The polarization response file (\texttt{.prsp}) is a multi-extension FITS file that stores the instrument's response to both polarized and unpolarized incident radiation as a function of input energy (E$_\mathrm{in}$), input polarization angle (PA$_\mathrm{in}$), and detected scattering angle (SA). The file contains five binary table extensions, the first three extensions store the information about the bin edges of the three dimensions of the response (energy, PA and scattering angle) and the last two extensions store the matrices.

The first extension \texttt{INEBOUNDS} lists the input energy bin edges in keV, the second extension \texttt{INPAVALS} defines the discrete input polarization angles in degrees used for $100\%$ polarized simulations, and the third extension \texttt{SABOUNDS} lists the scattering angle bins of the response.

The fourth extension, \texttt{SPECRESP POLMATRIX}, stores the response of the instrument to 100\% polarized incident photons using a 3D matrix of shape $m ({\rm E}_{\rm in}) \times n ({\rm PA}_{\rm in}) \times k ({\rm SA})$, where $m, n, k$ correspond to the number of input energies, input polarization angles and detected scattering bins respectively. The fifth extension, \texttt{SPECRESP UNPOLMATRIX}, stores the response of the instrument to unpolarized incident photons using a 2D matrix of shape $m ({\rm E_{in}}) \times k ({\rm SA})$ where $m$ and $k$ have the same definition as above. In both cases, the units of the matrix are cm$^\mathrm{2}$ indicating the effective area of each scattering bin for given input energy and polarization angle. This information is sufficient for \polpy to compute the response for any polarization fraction and for any input spectrum during the fitting process. Note that the response depends on the source direction relative to the IRF (see Section~\ref{sec:ref_frames}); hence, it has to be created for each source direction independently.

\subsection{Analysis Method}\label{sec:analysis_method}
The overall analysis methodology follows \polarpy, but with updates to the interpolation scheme and the likelihood formalism. The simulations are performed at $n$ PAs, but need to be evaluated at an arbitrary PA for fitting. For this we perform a harmonic interpolation between neighboring PA bins using the second-harmonic representation of the modulation pattern. This naturally accounts for the inherent \(180^\circ\) periodicity of linear polarization and provides a smooth interpolation of the detector response across the parameter space, as demonstrated in \citet{Bala2023}.

The geometry of \polar made it possible to undertake the analysis without binning the double-event photon pairs in angular bins; hence, \polarpy uses an unbinned Poisson likelihood for their calculations. However, this approach does not port well to instruments like \astrosat/CZTI or \daksha~(see Section~\ref{sec:daksha}) where we have only eight azimuthal bins, with non-uniform angular widths. Instead, we employ a binned Poisson likelihood, which naturally accounts for non-uniform binning in the modulation profiles. The log-likelihood is given by
\begin{equation}
    \ln \mathcal{L} = -\sum_{i=1}^{M}\left(N_i^{e}-N_i^{d}+N_i^{d}\ln\left[\frac{N_i^{d}}{N_i^{e}}\right]\right),
\end{equation}
where \(N_i^{d}\) and \(N_i^{e}\) respectively denote the detected and expected counts in the \(i\)-th SA bin, and \(M\) is the total number of bins. The expected counts are obtained by forward folding the spectro-polarimetric model through the instrument response.

\section{\polpy: Verification}~\label{sec:validation}
A critical step in developing \polpy was to verify the fitting process by testing the accuracy of the joint fit. Since there are no standard ``calibration'' sources when it comes to measuring the polarization of transient sources, we relied on injecting simulated sources in instrument mass models and recovering the injected parameters using \polpy. In the case of \astrosat/CZTI, the instrument collimators and other satellite components cause significant scattering, thereby weakening the polarization signature in the instrument and complicating both simulations and data analysis. Therefore, for this verification work, we used the mass model simulations of \polar~\citep{2017NIMPA.872...28K, 2018NIMPA.900....8L} and \daksha~\citep{Bhalerao2024, Bhalerao2024a}.

The goal here was to inject a GRB with known PA and PF into both instruments, then recover these values using \polpy. We first analyzed the data for individual instruments, followed by joint analysis for both instruments. We repeated this exercise for varied values of PA and PF. Below we present details of the injections and show the results obtained from the verification analysis.

\subsection{Injection}\label{sec:injection}
\begin{table}[t!]
\flushleft
\caption{Details of the injected GRB parameters. The source position in each instrument is given by the polar coordinates $\theta, \phi$.}
\label{tab:grb160325a_injection}
\begin{tabular}{lc}
\hline
Parameter & Value \\
\hline
Name & \thisgrb \\
RA & $16.003$\degr \\
Dec & $-72.048$\degr \\
UT & 2016-09-25T06:59:21 \\
$T_{90}$ & 42.94~s \\
Fluence\tablenotemark{a} & $1.98 \times 10^{-4}\mathrm{~erg~cm}^{-2}$\\
$\alpha$ & $-0.75$ \\
$\beta$ & $-2.32$ \\
$E_{\rm p}$ & $240$ keV \\
\daksha $\theta, \phi$ & $73.5$\degr, $-85.8$\degr\\
\polar $\theta, \phi$ & 76.4\degr, 236.2\degr \\
Injected PAs & 12\degr, 45\degr, 90\degr, 117\degr, 156\degr \\
Injected PFs & 30\%, 50\%, 80\% \\
\hline
\end{tabular}
\tablenotetext{a}{The spectral parameters and the fluence are from the \fermi/GBM 10-year catalog~\citep{fermi10yr}. The value of the fluence given here corresponds to the simulated GRB fluence as mentioned in the text; however the fluence of \thisgrborig was one tenth of this value.}
\end{table}
We picked one of the \astrosat/CZTI detected GRBs, \thisgrborig, as a source for our injection as we already had a well-defined spectral model for this GRB. We moved the detection epoch of this GRB such that we could use actual in-orbit background from \polar, making our simulations partly realistic. Since our goal was to validate the algorithms and their implementation in \polpy, we increased the fluence of the GRB by a factor of 10 as compared to the actual GRB, and refer to it as \thisgrb in this work for convenience. As a further simplifying assumption, we model the temporal profile of \thisgrb as a simple top-hat function with width equal to the \(T_{90}\) of the original burst. The instrument frame directions for \polar were calculated using the actual instrument data for the shifted epoch, while those for \daksha were calculated using assumed orbital parameters.

To validate the method across PA -- PF space, we picked five sky PAs: $[12^\circ, 45^\circ, 90^\circ, 117^\circ, 156^\circ]$, sampling the full range of linear polarization orientations; and three PF values: $[30\%, 50\%, 80\%]$ for injections. For each sky PA, the corresponding local PA was computed separately for \daksha and \polar which is needed to inject the GRB. To avoid any confusion and to be consistent in the succeeding discussions, the PA is always assumed to be in the sky, i.e. in the IAU frame, unless mentioned otherwise. Table~\ref{tab:grb160325a_injection} gives details about all the injection parameters. The following subsections briefly describe the instrument and injection details for \daksha and \polar.

\begin{figure}[t!]
        \centering
        \includegraphics[width=0.9\linewidth]{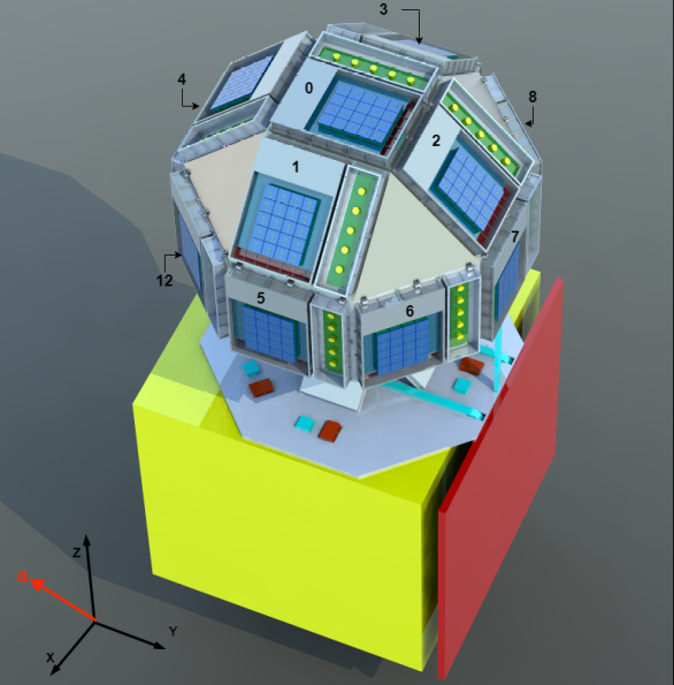}
        \caption{A rendering of \daksha showing the IRF (lower left), MEPs on the dome-like structure and their labels. Note that although the IRF axes are shown on lower-left, in reality the Z-axis of the IRF points through the MEP 0. The red arrow shows the approximate direction of the injected GRB.}
        \label{fig:daksha_face_name}
\end{figure}
\subsubsection{Daksha}\label{sec:daksha_details}
\daksha is a proposed all-sky high-energy transient mission consisting of two satellites orbiting the Earth on opposite sides~\citep{Bhalerao2024, Bhalerao2024a}. Each satellite has 17 Medium Energy Packages (MEPs), with 13 distributed along a dome-like structure (Figure~\ref{fig:daksha_face_name}) and four being located ``under'' the satellite bus, pointing in the $-Z$ direction. Each MEP consists of 20 CZT detectors identical to that of \astrosat/CZTI, enabling polarization measurements~\citep{Bala2023}. Each MEP on the dome points towards a different direction in the sky and acts as an independent detector.
\begin{figure*}
    \centering
    \includegraphics[width=\textwidth]{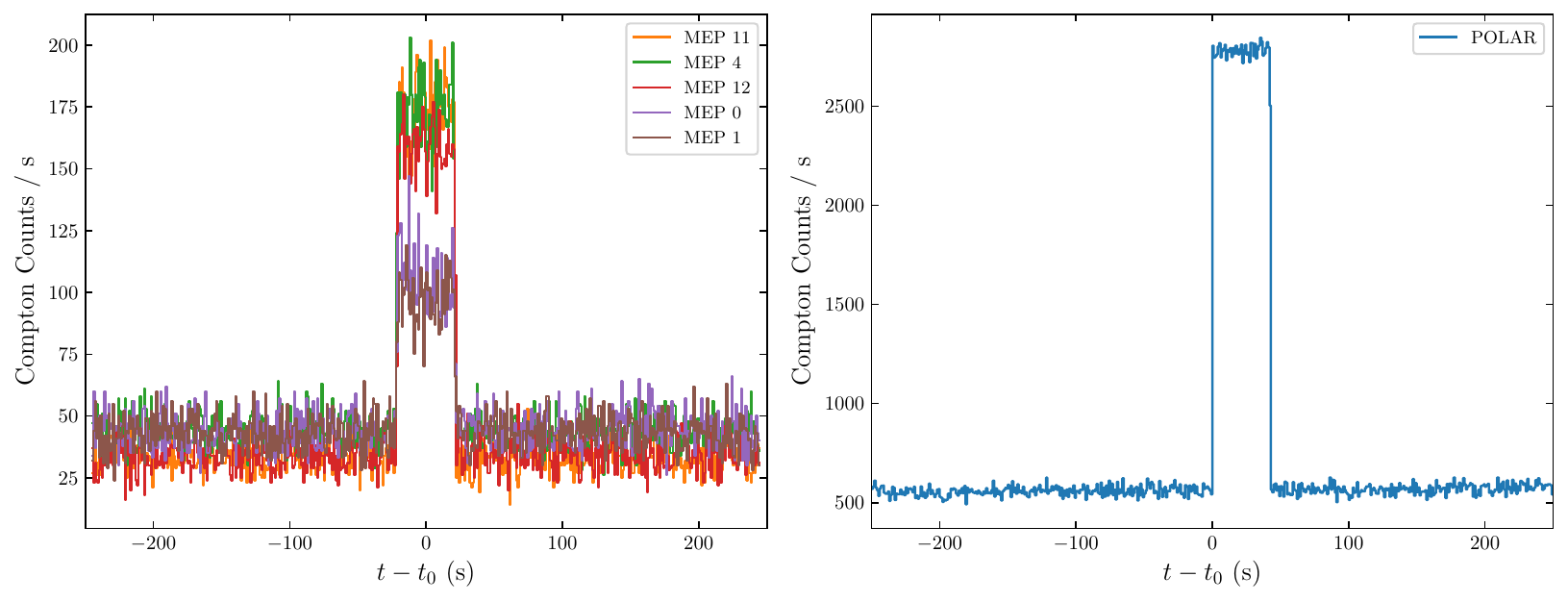}
    \caption{Compton event light curves showing a sample injected GRB in \daksha (left) and \polar (right). Note that the background events in \daksha are from simulations; however for \polar, real background event data around UTC 2016-09-25T06:59:21 is used.}\label{fig:grb_lc}
\end{figure*}

This makes \daksha an ideal choice for validating \polpy as with just one injection we can verify the basic code functionality by treating each MEP as an independent instrument observing the same GRB at different off-axis angles. For easier identification, we label the MEPs from 0 to 16 with respect to the \daksha IRF as shown in Figure~\ref{fig:daksha_face_name} and the discussion hereafter follows this labeling scheme when referring to MEPs. A typical GRB can be detected in up to eight MEPs in \daksha. \cite{Bala2023} showed that for polarization analysis, using data from the top five MEPs in terms of total detected counts is sufficient, and adding data from more MEPs does not significantly affect the measurements. Therefore, we use the same criteria for this work.
 
The selected GRB has $\theta,\phi = 73.5^\circ, -85.8^\circ$ in the \daksha IRF. Given this, it is closest to the MEP~11 surface normal, making it the top MEP with the largest number of counts, while the other four MEPs are 4, 12, 0, and 1. For injection, we use the \daksha mass model described in~\citep{Bala2023}. We simulate the GRB with given spectral shape, 100\% polarized photons and five PAs in the ILTP frame of \daksha corresponding to the sky PAs as given in Table~\ref{tab:grb160325a_injection}.
An additional simulation having the same spectral shape but unpolarized photons is also performed, and the final ``event-list" for a given PA -- PF injection is created by mixing appropriate fractions of photons from the unpolarized and  100\% polarized simulations. For background, we use the contributions from Cosmic X-ray Background (CXB) and Earth hard X-ray albedo in a similar way as described in~\cite{Bala2023}. The polarization response is created following the standards given in Section~\ref{sec:rsp_format} with PA bins of 10 degrees and a logarithmic input energy grid of 57 energies from 100 keV to 1000 keV. Figure~\ref{fig:grb_lc} shows the injected GRB light-curve for the top five MEPs.

\subsubsection{POLAR}\label{sec:polar_details}
\polar, launched on 15 September 2016, was a wide-field Compton polarimeter flown on board the Chinese Tiangong-2 space laboratory~\citep{Produit2018}. The detector consisted of 1600 plastic scintillator bars and was optimized for GRB polarimetry in the 50~--~500~keV energy range, with a large field of view of $\sim2\pi$~steradians. Polarization is measured through the azimuthal distribution of Compton-scattering angles reconstructed from coincident two-hit events in the detector array. During its approximately six months of operation, \polar detected 55 GRBs jointly with other instruments~\citep{2017ICRC...35..640X}, and polarization measurements of 14 GRBs were reported in~\citet{2019NatAs...3..258Z, Burgess2019, Kole2020}.

For \polar injections, we use the detailed simulation framework described in~\citet{2017NIMPA.872...28K,2018NIMPA.900....8L} to simulate the response to the source spectrum and polarization properties given in Table~\ref{tab:grb160325a_injection}. Since the actual \thisgrborig occurred before the launch of \polar, it was not observed by the instrument. We therefore use its sky position and spectral properties only as those of a representative GRB \thisgrb, while adopting a real \polar spacecraft attitude and in-orbit background data from an interval starting at 2016-09-25T06:59:21. For this attitude, the source is located at $(\theta,\phi)=(76.4^\circ,236.2^\circ)$ in the \polar IRF. The relatively large off-axis angle also provides a useful test of the coordinate transformations and the boundary treatment of the polarization response implemented in \polpy.

Using the source incidence angle in the \polar IRF, we generated the spectral and polarimetric responses at 150 monochromatic energies spanning 7.5--752.5~keV in 5~keV steps. For each energy, responses were simulated for 61 polarization angles from $0^\circ$ to $180^\circ$ in $3^\circ$ steps for a 100\% polarized source, together with one unpolarized case, giving a total grid of 9300 response matrices. For the GRB injection, photons following the adopted spectral shape are simulated at the local polarization angles corresponding to the five injected sky PAs listed in Table~\ref{tab:grb160325a_injection}. Source events corresponding to the required PFs are then constructed by mixing the polarized and unpolarized components. Source events are eventually combined with the selected real in-orbit \polar background, as shown in the right panel of Figure~\ref{fig:grb_lc}.

The resulting Compton events are reduced to their azimuthal scattering-angle distributions and converted into the common \polpy event format. The corresponding polarization response is constructed from the \polar mass-model simulations following the standardized response format described in Section~\ref{sec:rsp_format}. Unlike \daksha, for which each MEP provides eight azimuthal bins, the \polar analysis uses a finely sampled scattering-angle distribution with 360 bins. The different detector geometry, angular binning, and local reference frame of \polar provide an independent test of whether \polpy can recover a common sky-frame PA and PF. Subsequently, combining the \polar and \daksha injections provide the ultimate test of performing multi-instrument polarimetric fit using \polpy.

\subsection{Results}
\begin{figure*}[t!]
    \centering
    \includegraphics[width=\linewidth]{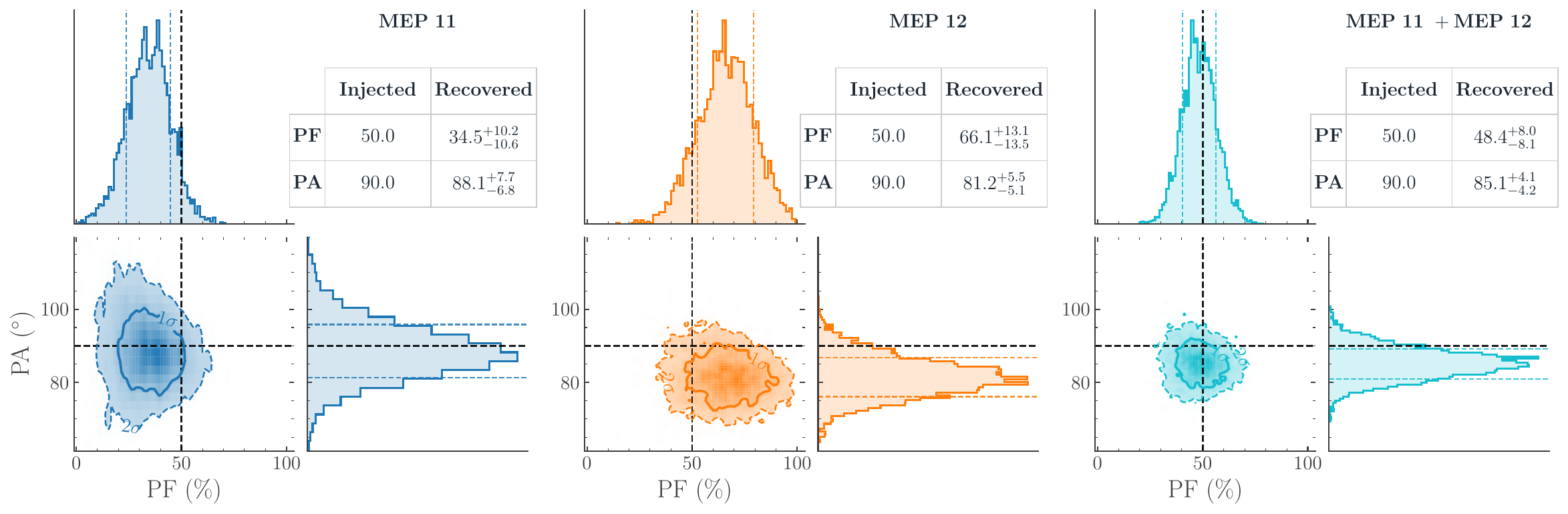}
    \caption{Polarization posterior distributions of single-MEP and two-MEP joint fit using \polpy for \daksha only injection case. The injected PA and PF are 90\degr and 50\% respectively. In each 2D contour panel, the inner solid line and outer dashed line trace the $1\sigma$ and $2\sigma$ enclosed confidence regions, respectively. Dashed black lines indicate the injected ground-truth parameters. Embedded summary tables report the injected parameters illustrating how combining statistics from two (or more) MEPs of \daksha tightens the recovered confidence interval.}
    \label{fig:daksha_one_two_face}
\end{figure*}
We present the results from our injection and validation analysis here. We first discuss the case of a joint fit between different \daksha MEPs as a single-instrument case (Section~\ref{sec:daksha}) and then discuss the results for the joint \daksha + \polar fit (Section~\ref{sec:jointpol}). In both cases, we do not perform a joint spectro-polarimetric fit, as the main purpose of this exercise was to verify the polarization fitting aspects of \polpy. Allowing for spectral fitting would have added additional degrees of freedom that are not related to polarization measurements, affecting the results and conclusions. However, to account for the statistical variations in our simulations, we do not freeze the spectral parameters completely but instead use ``truncated" priors for them, allowing them to vary around the mean injected values.
\subsubsection{Daksha Only}\label{sec:daksha}
\begin{figure*}
    \centering
    \includegraphics[width=1\linewidth]{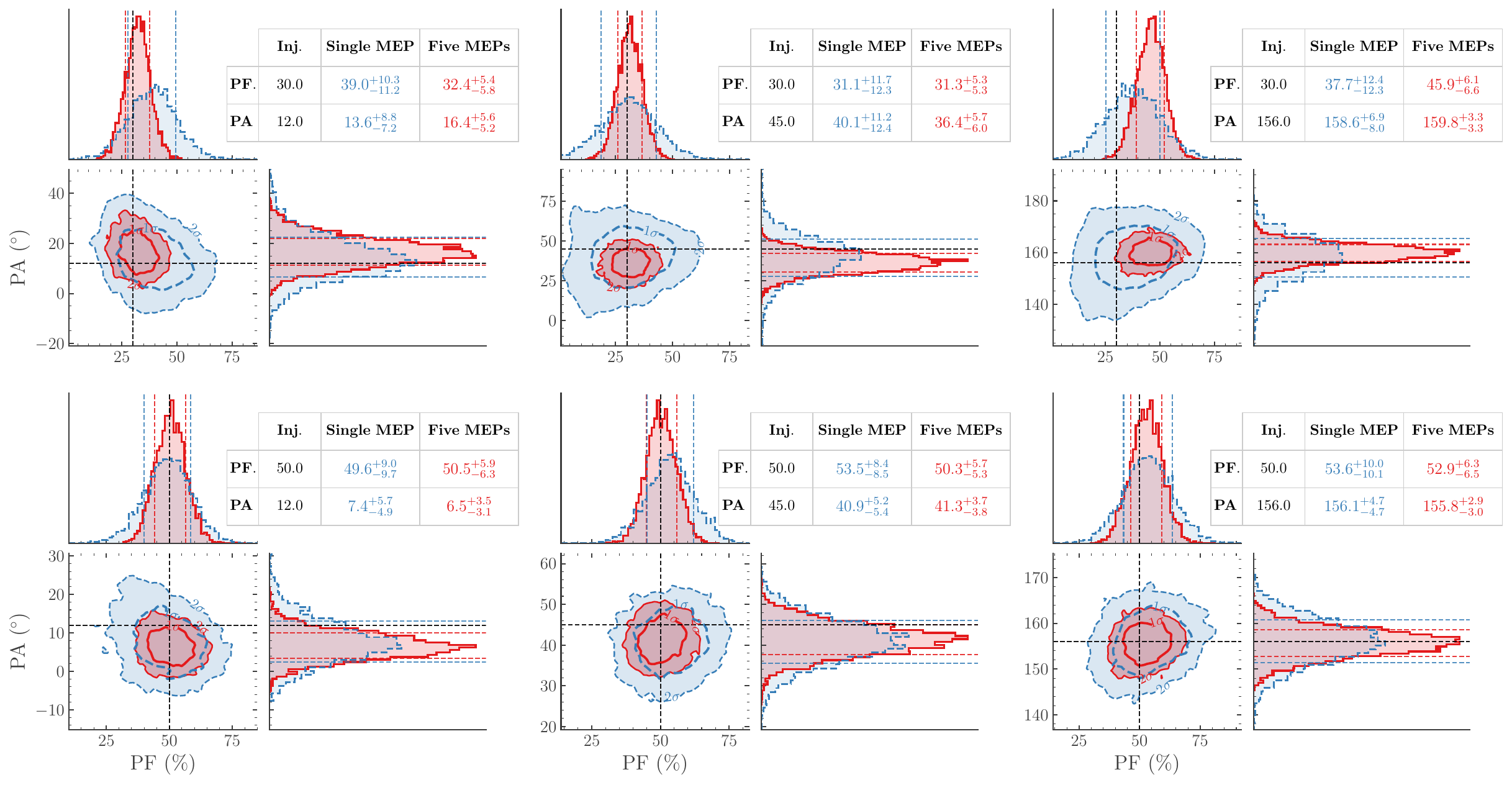}
\caption{Comparison of polarization posterior distributions for single-MEP and 5-MEP joint fit for ``\daksha-only'' injections. Parameter constraints are evaluated across two injected PFs and three PAs. Dashed blue contours and histograms represent the optimal single-detector MEP fit, while solid red distributions show the simultaneous 5-MEP joint recovery. Solid and dashed 2D contour lines mark the $1\sigma$ and $2\sigma$ confidence regions, and the dashed black lines indicate the injected input parameters.}
\label{fig:daksha_mult_face_fit}
\end{figure*}

Figure~\ref{fig:daksha_one_two_face} shows the inferred polarization parameters for the injected case of PA = 90\degr and PF = 50\% for the top two MEPs. It shows that, while the injected value remains within the $2\sigma$ contours of the ``only MEP 11'' and ``only MEP 12'' fits, the joint MEP 11 + MEP 12 fit leads to a much smaller and thus better-constrained contour. The result also shows that for an individual MEP as well as the joint-fit case, the PA is recovered correctly, validating one of the critical aspects of \polpy. Figure~\ref{fig:daksha_mult_face_fit} presents the injection--recovery results for a larger set: three simulated PAs and two PFs. The top MEP fit is consistent with the injected polarization parameters. However, the joint fit using the top five \daksha MEPs clearly yields tighter posterior distributions and recovered values. This demonstrates that there is a clear advantage of performing a joint multi-detector analysis, which significantly improves the constraints on both the polarization fraction and angle by combining the modulation patterns from different detector orientations. This result is consistent with previously published \daksha results where a multi-MEP fit was found to deliver superior recovery of injected values~\citep{Bala2023}.

\subsubsection{Daksha + POLAR}\label{sec:jointpol}
\begin{figure*}[!h]
    \centering
    \includegraphics[width= 1\linewidth]{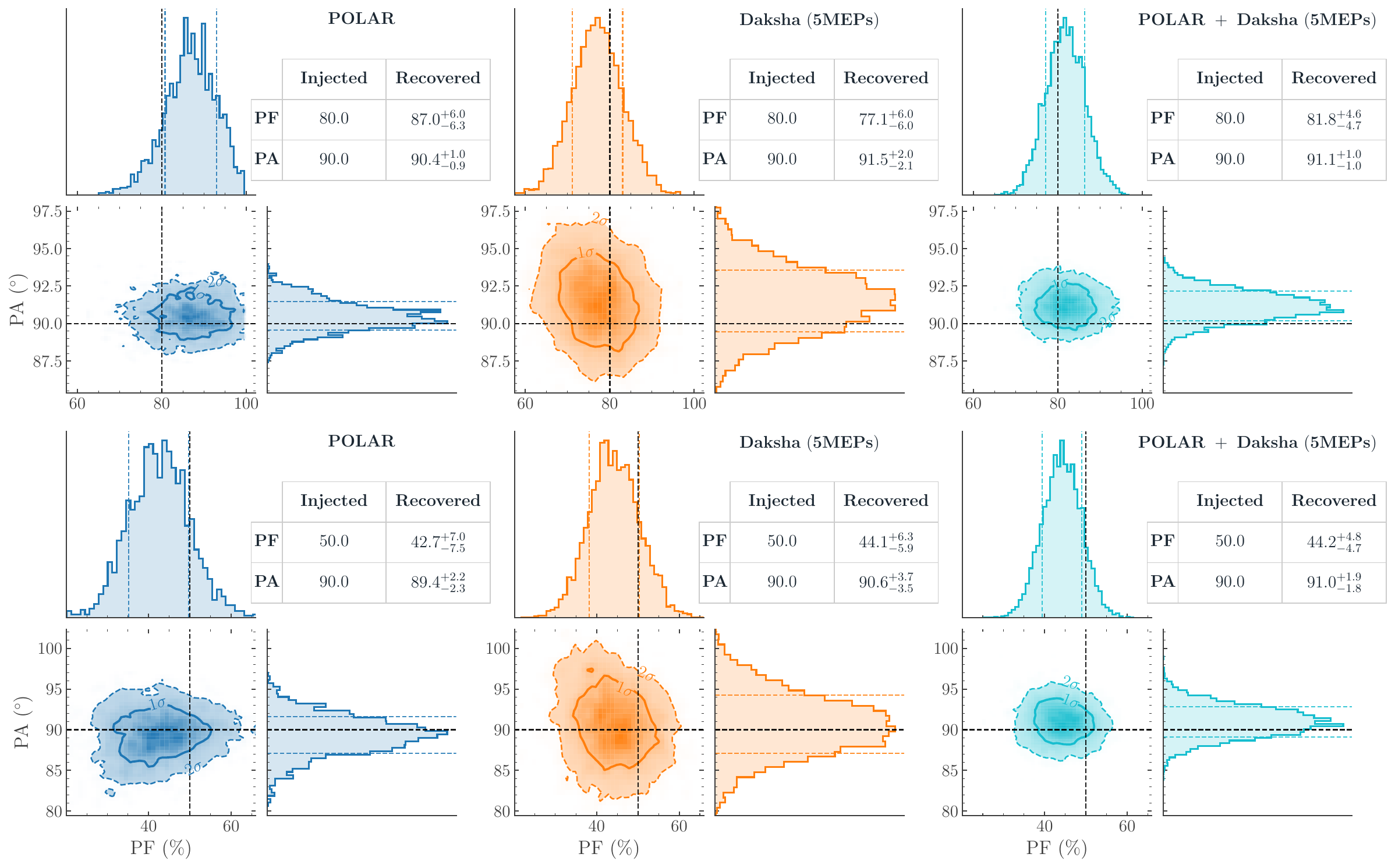}
    \caption{Comparison of polarization parameter recoveries for \polar, \daksha five-MEPs, and the joint \polar + \daksha fit for injected PA of 90\degr and two different PFs (Top Row: 80\%, Bottom Row: 50\%). It can be seen that for all three cases, \polpy recovers the injected values fairly well. The joint-fit posteriors also highlight the importance of cross-instrument fitting by clearly obtaining more constrained posteriors compared to single instrument fit only case.}
    \label{fig:polar_daksha_joint_comparison_combined}
\end{figure*}

\begin{figure*}[!hbt]
    \centering
    \includegraphics[width=1\linewidth]{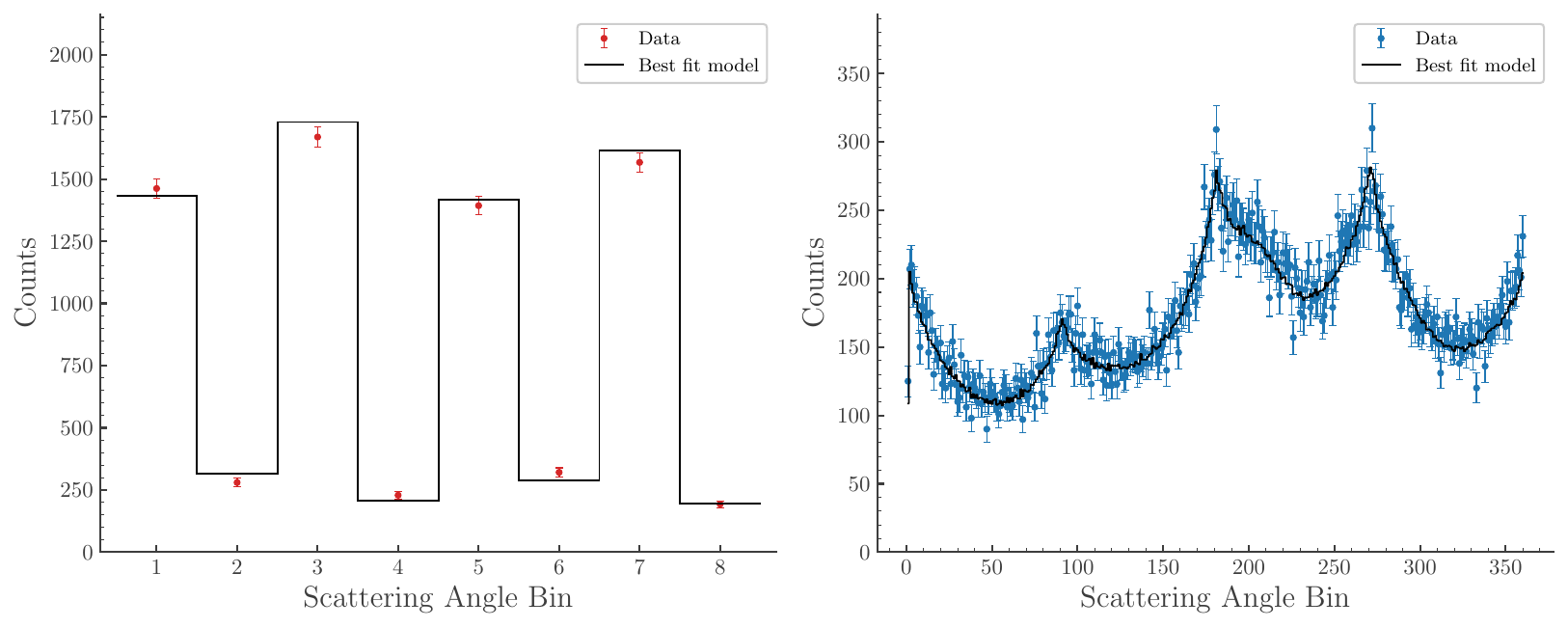}
    \caption{Best fit modulation curves from \daksha MEP 11 (left) and \polar (right) for the first joint fit case discussed in Figure~\ref{fig:polar_daksha_joint_comparison_combined}. Red and blue data points represent detected Compton counts in specified azimuthal scattering bins for \daksha and \polar, respectively. The grey line represents the best-fit model obtained from the joint fit.\label{fig:modulation_curves}}
\end{figure*}

Next we show the results of the joint fit between the \daksha and \polar simulations, the most complete case of verifying \polpy's functionality. Figure~\ref{fig:polar_daksha_joint_comparison_combined} shows the posteriors for individual instrument-level fits for \polar, \daksha (combined five MEP fit) and the joint \daksha{}--\polar fit for a case with an injected PA of $90^\circ$ and PF of $80\%$. Note that the GRB is at a large off-axis angle in the \polar frame, reducing its sensitivity to recover the polarization. However, the joint fit improves these results, giving tighter posteriors that are much closer to the true value. Additionally, the best-fit modulation curves for both instruments are shown in Figure~\ref{fig:modulation_curves} indicating very good match with the data for the best fit parameters.

\section{Discussion}~\label{sec:discussion}
The validation exercise summarized in Section~\ref{sec:validation} shows that \polpy correctly recovers polarization parameters for both \polar and \daksha. This is enabled by the uniform framework on which it relies (Section~\ref{sec:overview}). Specifically, fixing common reference frames (Section~\ref{sec:ref_frames}) removes the ambiguity while fitting the PA across instruments, while the standardized data (\texttt{.pevt}) and response (\texttt{.prsp}) formats (Section~\ref{sec:data_format}) allow for joint fits across any number of instruments. This is precisely the capability that was unavailable for the two GRBs co-detected by \polar and \astrosat/CZTI, where differences in the respective analysis pipelines prevented a joint fit that might have reconciled the discrepant polarization fractions reported by \citet{Kole2020} and \citet{Chattopadhyay2022}.

With the framework validated, the immediate application of \polpy is a joint analysis of the two GRBs co-detected by \astrosat/CZTI and \polar, which will provide the first direct test of whether the discrepancy between the two samples reflects an astrophysical difference, an instrumental systematic, or a pipeline-level effect (Pathak et al.\ 2026, in preparation). Looking further ahead, the same standardized data and response formats make it possible to uniformly process \astrosat/CZTI and \polar GRB data into a single, uniformly analyzed polarization catalog with \polpy, increasing the sample size available for population-level studies of GRB prompt-emission polarization (Pathak et al.\ 2027, in preparation). More generally, the mission-agnostic design of \polpy positions it to incorporate data from upcoming polarimeters such as \polarb, \cosi and \daksha as they come online, providing a common analysis backbone for GRB polarimetry through the next decade.

Plans to upgrade \polpy in the coming years are already under discussion. A near-term goal will be to make the necessary modifications to it and \threeml such that physical models can be directly folded to match the observed data~\citep[e.g.][]{Gill2026}. In addition, the prospect of moving the fitting to Stokes parameter space is also being considered; however, for off-axis polarimetry, defining Stokes parameters is not straightforward and extensive testing using simulations is necessary. Furthermore, it can be noted that, currently, \polpy only supports transient polarimetry and does not include a way to analyze data for persistent sources. This choice was made due to two main reasons: (a) the immediate need of a framework that can be used to fit the \polar and \astrosat/CZTI co-detections and (b) handling background for transient sources is significantly simpler than for persistent sources; hence, adding that capability would have slowed the development process. A long term goal will be to include this functionality in future releases.

\section{Conclusion}~\label{sec:conclusion}
With the renewed interest in X-ray/gamma-ray polarimetry over the past decade, a multitude of instruments have been proposed, launched, or are currently under development. However, the lack of standardized data formats and analysis pipelines across different missions has hindered joint analyses and population-level studies. In this article, we present \polpy, a common software for high-energy transient polarimetry based on standard response and data formats. We demonstrate its functionality by applying it to injected GRBs and recovering the injected parameters correctly for single-instrument as well as multi-instrument fit cases. In the coming years, \polpy will be upgraded to incorporate direct physical model fitting for polarization analysis, potentially explore the use of a Stokes parameter fit, and add functionality to analyze polarization data from persistent sources. By providing an open-source, standardized, and user-friendly framework, we hope \polpy will foster collaboration within the high-energy astrophysics community and maximize the scientific output of past, current, and future polarimetric observations.

\begin{acknowledgments}
We acknowledge support from the Swiss National Science Foundation through Scientific Exchanges grants 216920 and 230988, which supported the workshops that initiated this development. We thank the Space Program Office (SPO) of the Indian Space Research Organisation for its Announcement of Opportunity for space astrophysics missions, under which \daksha was proposed. Development of the \daksha Medium Energy Package laboratory model was started with funding support from SPO, and continued with support from all partner organisations. We thank the administrative and support staff at all partner institutes for their help in all \daksha-related matters.

Part of this work was supported by S.M.'s appointment to the NASA Postdoctoral Program at the NASA Goddard Space Flight Center, administered by Oak Ridge Associated Universities under contract with NASA.

We thank 3ML developer Niccolo Di Lalla from Stanford University for his valuable inputs and help regarding integrating \polpy with \threeml. S.M. and H. L. acknowledge use of High Performance Computing (HPC) facilities at Raman Research Institute and University of Geneva, respectively, for performing the simulations.
\end{acknowledgments}

\begin{contribution}
%%This section gives authors the space to recognize author contributions. The text inside this environment is NOT counted towards the total word quanta. At a minimum, manuscripts are expected to include this text:

All authors contributed equally to this collaborative work.

%% But authors are expected to provide more specific details, e.g. 
%%
%%SC was responsible for writing and submitting the manuscript.
%%WWM came up with the initial research concept and edited the manuscript.
%%OTS obtained the funding and edited the manuscript.
%%EBF provided the formal analysis and validation. He also edited the manuscript.
%%GEH Supervised the undergraduates, wrote the software and administers the project github and Zenodo repositories.
%%
%% Authors can use the Contributor Role Taxonomy (CRediT) at
%% for ideas on how write a good statement tailored to their needs.

\end{contribution}

%% \facilities{}

%% the manuscript. Authors should  list each code and include either a
\software{Astropy~\citep{2013A&A...558A..33A,2018AJ....156..123A,2022ApJ...935..167A}, Matplotlib~\citep{matplotlib2007}, NumPy~\citep{numpy2020}, SciPy~\citep{scipy2020}, \threeml~\citep{2015arXiv150708343V}}


% \appendix

% sample text

%% For this sample we use BibTeX plus aasjournalv7.bst to generate the

\bibliography{polpy}

@article{Weisskopf2022,
  author =        {Weisskopf, Martin C. and Soffitta, Paolo and
                   Baldini, Luca and Ramsey, Brian D. and
                   O'Dell, Stephen L. and Romani, Roger W. and
                   Matt, Giorgio and Deininger, William D. and
                   Baumgartner, Wayne H. and Bellazzini, Ronaldo and
                   Costa, Enrico and Kolodziejczak, Jeffery J. and
                   Latronico, Luca and Marshall, Herman L. and
                   Muleri, Fabio and Bongiorno, Stephen D. and
                   Tennant, Allyn and Bucciantini, Niccolo and
                   Dovciak, Michal and Marin, Fr{\'e}d{\'e}ric and
                   Marscher, Alan and Poutanen, Juri and Slane, Pat and
                   Turolla, Roberto and Kalinowski, William and
                   Marco, Alessandro Di and Fabiani, Sergio and
                   Minuti, Massimo and Monaca, Fabio La and
                   Pinchera, Michele and Rankin, John and
                   Sgr{\`o}, Carmelo and Trois, Alessio and Xie, Fei and
                   Alexander, Cheryl and Allen, D. Zachery and
                   Amici, Fabrizio and Andersen, Jason and
                   Antonelli, Angelo and Antoniak, Spencer and
                   Attin{\'a}, Primo and Barbanera, Mattia and
                   Bachetti, Matteo and Baggett, Randy M. and
                   Bladt, Jeff and Brez, Alessandro and
                   Bonino, Raffaella and Boree, Christopher and
                   Borotto, Fabio and Breeding, Shawn and
                   Brienza, Daniele and Bygott, H. Kyle and
                   Caporale, Ciro and Cardelli, Claudia and
                   Carpentiero, Rita and Castellano, Simone and
                   Castronuovo, Marco and Cavalli, Luca and
                   Cavazzuti, Elisabetta and Ceccanti, Marco and
                   Centrone, Mauro and Citraro, Saverio and
                   D'Amico, Fabio and D'Alba, Elisa and Gesu, Laura Di and
                   Monte, Ettore Del and Dietz, Kurtis L. and
                   Lalla, Niccol{\`o} Di and Persio, Giuseppe Di and
                   Dolan, David and Donnarumma, Immacolata and
                   Evangelista, Yuri and Ferrant, Kevin and
                   Ferrazzoli, Riccardo and Ferrie, MacKenzie and
                   Footdale, Joseph and Forsyth, Brent and
                   Foster, Michelle and Garelick, Benjamin and
                   Gunji, Shuichi and Gurnee, Eli and Head, Michael and
                   Hibbard, Grant and Johnson, Samantha and Kelly, Erik and
                   Kilaru, Kiranmayee and Lefevre, Carlo and
                   Roy, Shelley Le and Loffredo, Pasqualino and
                   Lorenzi, Paolo and Lucchesi, Leonardo and
                   Maddox, Tyler and Magazzu, Guido and Maldera, Simone and
                   Manfreda, Alberto and Mangraviti, Elio and
                   Marengo, Marco and Marrocchesi, Alessandra and
                   Massaro, Francesco and Mauger, David and
                   McCracken, Jeffery and McEachen, Michael and
                   Mize, Rondal and Mereu, Paolo and Mitchell, Scott and
                   Mitsuishi, Ikuyuki and Morbidini, Alfredo and
                   Mosti, Federico and Nasimi, Hikmat and Negri, Barbara and
                   Negro, Michela and Nguyen, Toan and Nitschke, Isaac and
                   Nuti, Alessio and Onizuka, Mitch and
                   Oppedisano, Chiara and Orsini, Leonardo and
                   Osborne, Darren and Pacheco, Richard and
                   Paggi, Alessandro and Painter, Will and
                   Pavelitz, Steven D. and Pentz, Christina and
                   Piazzolla, Raffaele and Perri, Matteo and
                   {Pesce-Rollins}, Melissa and Peterson, Colin and
                   Pilia, Maura and Profeti, Alessandro and
                   Puccetti, Simonetta and Ranganathan, Jaganathan and
                   Ratheesh, Ajay and Reedy, Lee and Root, Noah and
                   Rubini, Alda and Ruswick, Stephanie and
                   Sanchez, Javier and Sarra, Paolo and
                   Santoli, Francesco and Scalise, Emanuele and
                   Sciortino, Andrea and Schroeder, Christopher and
                   Seek, Tim and Sosdian, Kalie and Spandre, Gloria and
                   Speegle, Chet O. and Tamagawa, Toru and
                   Tardiola, Marcello and Tobia, Antonino and
                   Thomas, Nicholas E. and Valerie, Robert and
                   Vimercati, Marco and Walden, Amy L. and
                   Weddendorf, Bruce and Wedmore, Jeffrey and
                   Welch, David and Zanetti, Davide and
                   Zanetti, Francesco},
  journal =       {JATIS},
  month =         apr,
  number =        {2},
  pages =         {026002},
  publisher =     {SPIE},
  title =         {Imaging {{X-ray Polarimetry Explorer}}: Prelaunch},
  volume =        {8},
  year =          {2022},
  doi =           {10.1117/1.JATIS.8.2.026002},
  issn =          {2329-4124, 2329-4221},
}

@article{Produit2018,
  author =        {Produit, N. and Bao, T. W. and Batsch, T. and
                   Bernasconi, T. and Britvich, I. and Cadoux, F. and
                   Cernuda, I. and Chai, J. Y. and Dong, Y. W. and
                   Gauvin, N. and Hajdas, W. and Kole, M. and
                   Kong, M. N. and Kramert, R. and Li, L. and Liu, J. T. and
                   Liu, X. and Marcinkowski, R. and Orsi, S. and
                   Pohl, M. and Rapin, D. and Rybka, D. and
                   Rutczynska, A. and Shi, H. L. and Socha, P. and
                   Sun, J. C. and Song, L. M. and Szabelski, J. and
                   Traseira, I. and Xiao, H. L. and Wang, R. J. and
                   Wen, X. and Wu, B. B. and Zhang, L. and Zhang, L. Y. and
                   Zhang, S. N. and Zhang, Y. J. and Zwolinska, A.},
  journal =       {Nuclear Instruments and Methods in Physics Research
                   Section A: Accelerators, Spectrometers, Detectors and
                   Associated Equipment},
  month =         jan,
  pages =         {259--268},
  title =         {Design and Construction of the {{POLAR}} Detector},
  volume =        {877},
  year =          {2018},
  doi =           {10.1016/j.nima.2017.09.053},
  issn =          {0168-9002},
}

@article{Kole2020,
  author =        {{Kole, M.} and {De Angelis, N.} and {Berlato, F.} and
                   {Burgess, J. M.} and {Gauvin, N.} and {Greiner, J.} and
                   {Hajdas, W.} and {Li, H. C.} and {Li, Z. H.} and
                   {Pollo, A.} and {Produit, N.} and {Rybka, D.} and
                   {Song, L. M.} and {Sun, J. C.} and {Szabelski, J.} and
                   {Tymieniecka, T.} and {Wang, Y. H.} and {Wu, B. B.} and
                   {Wu, X.} and {Xiong, S. L.} and {Zhang, S. N.} and
                   {Zhang, Y. J.}},
  journal =       {A\&A},
  pages =         {A124},
  title =         {The POLAR gamma-ray burst polarization catalog},
  volume =        {644},
  year =          {2020},
  doi =           {10.1051/0004-6361/202037915},
  url =           {https://doi.org/10.1051/0004-6361/202037915},
}

@article{Bhalerao2017a,
  author =        {Bhalerao, V. and Bhattacharya, D. and Vibhute, A. and
                   Pawar, P. and Rao, A. R. and Hingar, M. K. and
                   Khanna, Rakesh and Kutty, A. P.K. and Malkar, J. P. and
                   Patil, M. H. and Arora, Y. K. and Sinha, S. and
                   Priya, P. and Samuel, Essy and Sreekumar, S. and
                   Vinod, P. and Mithun, N. P.S. and Vadawale, S. V. and
                   Vagshette, N. and Navalgund, K. H. and Sarma, K. S. and
                   Pandiyan, R. and Seetha, S. and Subbarao, K.},
  journal =       {Journal of Astrophysics and Astronomy},
  month =         jun,
  number =        {2},
  pages =         {1--10},
  publisher =     {Springer India},
  title =         {The {{Cadmium Zinc Telluride Imager}} on
                   {{AstroSat}}},
  volume =        {38},
  year =          {2017},
  doi =           {10.1007/s12036-017-9447-8},
  issn =          {09737758},
}

@article{Chattopadhyay2022,
  author =        {Chattopadhyay, Tanmoy and Gupta, Soumya and
                   Iyyani, Shabnam and Saraogi, Divita and
                   Sharma, Vidushi and Tsvetkova, Anastasia and
                   Ratheesh, Ajay and Gupta, Rahul and Mithun, N. P. S. and
                   Vaishnava, C. S. and Prasad, Vipul and Aarthy, E. and
                   Kumar, Abhay and Rao, A. R. and Vadawale, Santosh and
                   Bhalerao, Varun and Bhattacharya, Dipankar and
                   Vibhute, Ajay and Frederiks, Dmitry},
  journal =       {The Astrophysical Journal},
  month =         aug,
  number =        {1},
  pages =         {12},
  publisher =     {IOP Publishing},
  title =         {Hard {{X-Ray Polarization Catalog}} for a {{Five-year
                   Sample}} of {{Gamma-Ray Bursts Using AstroSat CZT
                   Imager}}},
  volume =        {936},
  year =          {2022},
  doi =           {10.3847/1538-4357/ac82ef},
  issn =          {0004-637X},
}

@article{Dean2008,
  author =        {Dean, A. J. and Clark, D. J. and Stephen, J. B. and
                   McBride, V. A. and Bassani, L. and Bazzano, A. and
                   Bird, A. J. and Hill, A. B. and Shaw, S. E. and
                   Ubertini, P.},
  journal =       {Science},
  month =         aug,
  number =        {5893},
  pages =         {1183--1185},
  publisher =     {American Association for the Advancement of Science},
  title =         {Polarized {{Gamma-Ray Emission}} from the {{Crab}}},
  volume =        {321},
  year =          {2008},
  doi =           {10.1126/science.1149056},
}

@article{Gotz2009,
  author =        {G{\"o}tz, Diego and Laurent, Philippe and
                   Lebrun, Fran{\c c}ois and Daigne, Fr{\'e}d{\'e}ric and
                   Bonjak, Eljka},
  journal =       {Astrophysical Journal},
  month =         mar,
  number =        {2 PART 2},
  pages =         {L208-L212},
  title =         {Variable Polarization Measured in the Prompt Emission
                   of {{GRB}} 041219a Using Ibis on Board Integral},
  volume =        {695},
  year =          {2009},
  doi =           {10.1088/0004-637X/695/2/L208},
  issn =          {15384357},
}

@article{Laurent2011,
  author =        {Laurent, P. and Rodriguez, J. and Wilms, J. and
                   Cadolle Bel, M. and Pottschmidt, K. and Grinberg, V.},
  journal =       {Science},
  month =         apr,
  number =        {6028},
  pages =         {438--439},
  publisher =     {American Association for the Advancement of Science},
  title =         {Polarized {{Gamma-Ray Emission}} from the {{Galactic
                   Black Hole Cygnus X-1}}},
  volume =        {332},
  year =          {2011},
  doi =           {10.1126/science.1200848},
}

@incollection{Tomsick2024,
  author =        {Tomsick, John and Boggs, Steven and Zoglauer, Andreas and
                   Hartmann, Dieter H. and Ajello, Marco and Burns, Eric and
                   Fryer, Chris and Karwin, Chris and Kierans, Carolyn and
                   Lowell, Alex and Malzac, Julien and Roberts, Jarred and
                   {Saint-Hilaire}, Pascal and Shih, Albert and
                   Siegert, Thomas and Sleator, Clio and
                   Takahashi, Tadayuki and Tavecchio, Fabrizio and
                   Wulf, Eric and Beechert, Jacqueline and
                   Gulick, Hannah and Joens, Alyson and Lazar, Hadar and
                   Neights, Eliza and Martinez Oliveros, Juan Carlos and
                   Matsumoto, Shigeki and Melia, Tom and Yoneda, Hiroki and
                   Amman, Mark and Bal, Dhruv and {von Ballmoos}, Peter and
                   Bates, Hugh and B{\"o}ttcher, Markus and
                   Bulgarelli, Andrea and Cavazzuti, Elisabetta and
                   Chang, Hsiang-Kuang and Chen, Claire and Chu, Che-Yen and
                   Ciabattoni, Alex and Costamante, Luigi and
                   Dreyer, Lente and Fioretti, Valentina and
                   Fenu, Francesco and Gallego, Savitri and
                   Ghirlanda, Giancarlo and Grove, Eric and
                   Huang, Chien-You and Jean, Pierre and Khatiya, Nikita and
                   Kn{\"o}dlseder, J{\"u}rgen and Kraus, Martin and
                   Leising, Mark and Lewis, Tiffany and Lommler, Jan and
                   Marcotulli, Lea and Martinez Castellanos, Israel and
                   Mittal, Saurabh and Negro, Michela and
                   Al Nussirat, Samer and Nakazawa, Kazuhiro and
                   Oberlack, Uwe and Palmore, David and
                   Panebianco, Gabriele and Parmiggiani, Nicol{\`o} and
                   Pike, Sean and Rogers, Field and Schutte, Hester and
                   Sheng, Yong and Smale, Alan and Smith, Jacob Russell and
                   Trigg, Aaron and Venters, Tonia and Watanabe, Yu and
                   Zhang, Haocheng},
  booktitle =     {Proceedings of 38th {{International Cosmic Ray
                   Conference}} --- {{PoS}}({{ICRC2023}})},
  month =         sep,
  pages =         {745},
  publisher =     {SISSA Medialab},
  title =         {The {{Compton Spectrometer}} and {{Imager}}},
  volume =        {444},
  year =          {2024},
  doi =           {10.22323/1.444.0745},
}

@article{Kole2025,
  author =        {Kole, Merlin and De Angelis, Nicolas and He, Jiang and
                   Liu, Hongbang and Sun, Jianchao and Xie, Fei and
                   Zaid, Jimmy and Kole, Merlin and Angelis, Nicolas De and
                   He, Jiang and Liu, Hongbang and Sun, Jianchao and
                   Xie, Fei and Zaid, Jimmy},
  journal =       {SpaceTech},
  month =         dec,
  number =        {1},
  pages =         {2},
  publisher =     {Scilight Press},
  title =         {Design and {{Scientific Prospects}} of the {{POLAR-2
                   Mission}}},
  volume =        {1},
  year =          {2025},
}

@article{Bhalerao2024,
  author =        {Bhalerao, Varun and Vadawale, Santosh and
                   Tendulkar, Shriharsh and Bhattacharya, Dipankar and
                   Rana, Vikram and Adalja, Hitesh Kumar L. and
                   Belatikar, Hrishikesh and Bhaganagare, Mahesh and
                   Dewangan, Gulab and Ghodgaonkar, Abhijeet and
                   Kumar Goyal, Shiv and Gunasekaran, Suresh and
                   P J, Guruprasad and Koyande, Jayprakash G. and
                   Kulkarni, Salil and Kutty, {\relax APK} and
                   Ladiya, Tinkal and Mahapatra, Suddhasatta and
                   Marla, Deepak and Mate, Sujay and Mithun, N.P.S. and
                   Mote, Rakesh and Narang, Sanjoli and Nema, Ayush and
                   Nimbalkar, Sudhanshu and Pai, Archana and
                   Palit, Sourav and Patel, Arpit and Patel, Jinaykumar and
                   Pradeep, Priya and Ramachandran, Prabhu and
                   Bharath Saiguhan, B.S. and Saraogi, Divita and
                   Sawant, Disha and Shanmugam, M. and Sharma, Piyush and
                   Shetye, Amit and Singh, Nishant and Singh, Shreeya and
                   Singhal, Akshat and Sreekumar, S. and
                   Sridhar, Srividhya and Srinivasan, Rahul and
                   Tallur, Siddharth and Tiwari, Neeraj K. and
                   Lakshmi Vadladi, Amrutha and Vaishnava, C. S. and
                   Vishwakarma, Sandeep and Waratkar, Gaurav},
  journal =       {Exp Astron},
  month =         jun,
  number =        {3},
  pages =         {24},
  title =         {Daksha: On Alert for High Energy Transients},
  volume =        {57},
  year =          {2024},
  doi =           {10.1007/s10686-024-09926-y},
  issn =          {1572-9508},
}

@article{Bhalerao2024a,
  author =        {Bhalerao, Varun and Sawant, Disha and Pai, Archana and
                   Tendulkar, Shriharsh and Vadawale, Santosh and
                   Bhattacharya, Dipankar and Rana, Vikram and
                   Adalja, Hitesh Kumar L. and Anupama, G. C. and
                   Bala, Suman and Banerjee, Smaranika and
                   Basu, Judhajeet and Belatikar, Hrishikesh and
                   Beniamini, Paz and Bhaganagare, Mahesh and
                   Bhaskar, Ankush and Bhattacharjee, Soumyadeep and
                   Bose, Sukanta and Cenko, Brad and Vijay Chanda, Mehul and
                   Dewangan, Gulab and Dixit, Vishal and Dutta, Anirban and
                   Gawade, Priyanka and Ghodgaonkar, Abhijeet and
                   Kumar Goyal, Shiv and Gunasekaran, Suresh and
                   Hemanth, Manikantan and Hotokezaka, Kenta and
                   Iyyani, Shabnam and Guruprasad, P. J. and
                   Kasliwal, Mansi and G. Koyande, Jayprakash and
                   Kulkarni, Salil and Kutty, {\relax APK} and
                   Ladiya, Tinkal and Mahapatra, Suddhasatta and
                   Marla, Deepak and Mate, Sujay and Mehla, Advait and
                   Mithun, N. P. S. and More, Surhud and Mote, Rakesh and
                   Mukherjee, Dipanjan and Narang, Sanjoli and
                   Narendranath, Shyama and Nema, Ayush and
                   Nimbalkar, Sudhanshu and Nissanke, Samaya and
                   Palit, Sourav and Patel, Jinaykumar and Patel, Arpit and
                   Paul, Biswajit and Pradeep, Priya and
                   Ramachandran, Prabhu and Roy, Kinjal and
                   Saiguhan, B.S. Bharath and Saji, Joseph and
                   Saleem, M. and Saraogi, Divita and Sastry, Parth and
                   Shanmugam, M. and Sharma, Piyush and Shetye, Amit and
                   Singh, Nishant and Singh, Shreeya and Singhal, Akshat and
                   Sreekumar, S. and Sridhar, Srividhya and
                   Srinivasan, Rahul and Tallur, Siddharth and
                   K. Tiwari, Neeraj and Lakshmi Vadladi, Amrutha and
                   Vaishnava, C.S. and Vishwakarma, Sandeep and
                   Waratkar, Gaurav},
  journal =       {Exp Astron},
  month =         jun,
  number =        {3},
  pages =         {23},
  title =         {Science with the {{Daksha}} High Energy Transients
                   Mission},
  volume =        {57},
  year =          {2024},
  doi =           {10.1007/s10686-024-09923-1},
  issn =          {1572-9508},
}

@article{Bala2023,
  author =        {Bala, Suman and Mate, Sujay and Mehla, Advait and
                   Sastry, Parth and Mithun, N. P. S. and Palit, Sourav and
                   Chanda, Mehul Vijay and Saraogi, Divita and
                   Vaishnava, C. S. and Waratkar, Gaurav and
                   Bhalerao, Varun and Bhattacharya, Dipankar and
                   Tendulkar, Shriharsh and Vadawale, Santosh},
  journal =       {JATIS},
  month =         oct,
  number =        {4},
  pages =         {048002},
  publisher =     {SPIE},
  title =         {Prospects of Measuring Gamma-Ray Burst Polarization
                   with the {{Daksha}} Mission},
  volume =        {9},
  year =          {2023},
  doi =           {10.1117/1.JATIS.9.4.048002},
  issn =          {2329-4124, 2329-4221},
}

@article{Toma2009,
  author =        {Toma, Kenji and Sakamoto, Takanori and Zhang, Bing and
                   Hill, Joanne E. and McConnell, Mark L. and
                   Bloser, Peter F. and Yamazaki, Ryo and Ioka, Kunihito and
                   Nakamura, Takashi},
  journal =       {ApJ},
  month =         jun,
  number =        {2},
  pages =         {1042--1053},
  title =         {{{STATISTICAL PROPERTIES OF GAMMA-RAY BURST
                   POLARIZATION}}},
  volume =        {698},
  year =          {2009},
  doi =           {10.1088/0004-637X/698/2/1042},
  issn =          {0004-637X, 1538-4357},
}

@incollection{Vianello2016,
  author =        {Vianello, Giacomo and Lauer, Robert and
                   Younk, Patrick and Tibaldo, Luigi and
                   Burgess, James Michael and Ayala Solares, Hugo and
                   Harding, J. Patrick and Hui, C. Michelle and
                   Omodei, Nicola and Zhou, Hao},
  booktitle =     {Proceedings of {{The}} 34th {{International Cosmic
                   Ray Conference}} --- {{PoS}}({{ICRC2015}})},
  month =         aug,
  pages =         {1042},
  publisher =     {SISSA Medialab},
  title =         {The {{Multi-Mission Maximum Likelihood}} Framework},
  volume =        {236},
  year =          {2016},
  doi =           {10.22323/1.236.1042},
}

@article{Burgess2019,
  author =        {Burgess, J. M. and Kole, M. and Berlato, F. and
                   Greiner, J. and Vianello, G. and Produit, N. and
                   Li, Z. H. and Sun, J. C.},
  journal =       {A\&A},
  month =         jul,
  pages =         {A105},
  publisher =     {EDP Sciences},
  title =         {Time-Resolved {{GRB}} Polarization with {{POLAR}} and
                   {{GBM}} - {{Simultaneous}} Spectral and Polarization
                   Analysis with Synchrotron Emission},
  volume =        {627},
  year =          {2019},
  doi =           {10.1051/0004-6361/201935056},
  issn =          {0004-6361, 1432-0746},
}

@article{Chattopadhyay2014,
  author =        {Chattopadhyay, T. and Vadawale, S. V. and Rao, A. R. and
                   Sreekumar, S. and Bhattacharya, D.},
  journal =       {Experimental Astronomy},
  number =        {3},
  pages =         {555--577},
  title =         {Prospects of Hard {{X-ray}} Polarimetry with
                   {{Astrosat-CZTI}}},
  volume =        {37},
  year =          {2014},
  doi =           {10.1007/s10686-014-9386-1},
  issn =          {09226435},
}

@article{2015A&A...578A..73V,
  author =        {{Vadawale}, S.~V. and {Chattopadhyay}, T. and
                   {Rao}, A.~R. and {Bhattacharya}, D. and
                   {Bhalerao}, V.~B. and {Vagshette}, N. and {Pawar}, P. and
                   {Sreekumar}, S.},
  journal =       {\aap},
  month =         jun,
  pages =         {A73},
  title =         {{Hard X-ray polarimetry with Astrosat-CZTI}},
  volume =        {578},
  year =          {2015},
  doi =           {10.1051/0004-6361/201525686},
  eid =           {A73},
}

@phdthesis{aarthy2021multiwavelength,
  author =        {Aarthy, E and others},
  school =        {Indian Institute of Technology Gandhinagar},
  title =         {Multiwavelength polarimetry of astrophysical sources},
  year =          {2021},
}

@article{2022MNRAS.512.2827L,
  author =        {{Li}, Han-Cheng and {Produit}, Nicolas and
                   {Zhang}, Shuang-Nan and {Kole}, Merlin and
                   {Sun}, Jian-Chao and {Ge}, Ming-Yu and
                   {De Angelis}, Nicolas and {Hulsman}, Johannes and
                   {Li}, Zheng-Heng and {Song}, Li-Ming and
                   {Tymieniecka}, Teresa and {Wu}, Bo-Bing and {Wu}, Xin and
                   {Wang}, Yuan-Hao and {Xiong}, Shao-Lin and
                   {Zhang}, Yong-Jie and {Zhao}, Yi and
                   {Zheng}, Shi-Jie},
  journal =       {\mnras},
  month =         may,
  number =        {2},
  pages =         {2827-2840},
  title =         {{Gamma-ray polarimetry of the Crab pulsar observed by
                   POLAR}},
  volume =        {512},
  year =          {2022},
  doi =           {10.1093/mnras/stac522},
}

@techreport{george1992calibration,
  author =        {George, Ian M. and Zellar, Ron and Pence, William},
  institution =   {NASA High Energy Astrophysics Science Archive
                   Research Center (HEASARC)},
  number =        {CAL/GEN/92-002},
  type =          {OGIP Calibration Memo},
  title =         {The Calibration Requirements for Spectral Analysis},
  year =          {1992},
  url =           {https://heasarc.gsfc.nasa.gov/docs/heasarc/caldb/docs/memos/
                  cal_gen_92_002a/cal_gen_92_002a.pdf},
}

@article{2017NIMPA.872...28K,
  author =        {{Kole}, M. and {Li}, Z.~H. and {Produit}, N. and
                   {Tymieniecka}, T. and {Zhang}, J. and {Zwolinska}, A. and
                   {Bao}, T.~W. and {Bernasconi}, T. and {Cadoux}, F. and
                   {Feng}, M.~Z. and {Gauvin}, N. and {Hajdas}, W. and
                   {Kong}, S.~W. and {Li}, H.~C. and {Li}, L. and
                   {Liu}, X. and {Marcinkowski}, R. and {Orsi}, S. and
                   {Pohl}, M. and {Rybka}, D. and {Sun}, J.~C. and
                   {Song}, L.~M. and {Szabelski}, J. and {Wang}, R.~J. and
                   {Wang}, Y.~H. and {Wen}, X. and {Wu}, B.~B. and
                   {Wu}, X. and {Xiao}, H.~L. and {Xiong}, S.~L. and
                   {Zhang}, L. and {Zhang}, L.~Y. and {Zhang}, S.~N. and
                   {Zhang}, X.~F. and {Zhang}, Y.~J. and {Zhao}, Y.},
  journal =       {Nuclear Instruments and Methods in Physics Research
                   A},
  month =         nov,
  pages =         {28-40},
  title =         {{Instrument performance and simulation verification
                   of the POLAR detector}},
  volume =        {872},
  year =          {2017},
  doi =           {10.1016/j.nima.2017.07.070},
}

@article{2018NIMPA.900....8L,
  author =        {{Li}, Zhengheng and {Kole}, Merlin and
                   {Sun}, Jianchao and {Song}, Liming and
                   {Produit}, Nicolas and {Wu}, Bobing and
                   {Bao}, Tianwei and {Bernasconi}, Tancredi and
                   {Cadoux}, Franck and {Dong}, Yongwei and
                   {Feng}, Minzi and {Gauvin}, Neal and {Hajdas}, Wojtek and
                   {Li}, Hancheng and {Li}, Lu and {Liu}, Xin and
                   {Marcinkowski}, Radoslaw and {Pohl}, Martin and
                   {Rybka}, Dominik K. and {Shi}, Haoli and
                   {Szabelski}, Jacek and {Tymieniecka}, Teresa and
                   {Wang}, Ruijie and {Wang}, Yuanhao and {Wen}, Xing and
                   {Wu}, Xin and {Xiong}, Shaolin and {Zwolinska}, Anna and
                   {Zhang}, Li and {Zhang}, Laiyu and {Zhang}, Shuangnan and
                   {Zhang}, Yongjie and {Zhao}, Yi},
  journal =       {Nuclear Instruments and Methods in Physics Research
                   A},
  month =         aug,
  pages =         {8-24},
  title =         {{In-orbit instrument performance study and
                   calibration for POLAR polarization measurements}},
  volume =        {900},
  year =          {2018},
  doi =           {10.1016/j.nima.2018.05.041},
}

@inproceedings{2017ICRC...35..640X,
  author =        {{Xiong}, S. and {Wang}, Y. and {Li}, Z. and {Sun}, J. and
                   {Zhao}, Y. and {Li}, H. and {Huang}, Y. and
                   {Polar Collaboration}},
  booktitle =     {35th International Cosmic Ray Conference (ICRC2017)},
  month =         jul,
  pages =         {640},
  series =        {International Cosmic Ray Conference},
  title =         {{Overview of the GRB observation by POLAR}},
  volume =        {301},
  year =          {2017},
  doi =           {10.22323/1.301.0640},
  eid =           {640},
}

@article{2013A&A...558A..33A,
  author =        {{Astropy Collaboration} and {Robitaille}, Thomas P. and
                   {Tollerud}, Erik J. and {Greenfield}, Perry and
                   {Droettboom}, Michael and {Bray}, Erik and
                   {Aldcroft}, Tom and {Davis}, Matt and
                   {Ginsburg}, Adam and {Price-Whelan}, Adrian M. and
                   {Kerzendorf}, Wolfgang E. and {Conley}, Alexander and
                   {Crighton}, Neil and {Barbary}, Kyle and
                   {Muna}, Demitri and {Ferguson}, Henry and
                   {Grollier}, Fr{\'e}d{\'e}ric and {Parikh}, Madhura M. and
                   {Nair}, Prasanth H. and {Unther}, Hans M. and
                   {Deil}, Christoph and {Woillez}, Julien and
                   {Conseil}, Simon and {Kramer}, Roban and
                   {Turner}, James E.~H. and {Singer}, Leo and
                   {Fox}, Ryan and {Weaver}, Benjamin A. and
                   {Zabalza}, Victor and {Edwards}, Zachary I. and
                   {Azalee Bostroem}, K. and {Burke}, D.~J. and
                   {Casey}, Andrew R. and {Crawford}, Steven M. and
                   {Dencheva}, Nadia and {Ely}, Justin and
                   {Jenness}, Tim and {Labrie}, Kathleen and
                   {Lim}, Pey Lian and {Pierfederici}, Francesco and
                   {Pontzen}, Andrew and {Ptak}, Andy and
                   {Refsdal}, Brian and {Servillat}, Mathieu and
                   {Streicher}, Ole},
  journal =       {\aap},
  month =         {Oct},
  pages =         {A33},
  title =         {{Astropy: A community Python package for astronomy}},
  volume =        {558},
  year =          {2013},
  doi =           {10.1051/0004-6361/201322068},
  eid =           {A33},
}

@article{2018AJ....156..123A,
  author =        {{Astropy Collaboration} and {Price-Whelan}, A.~M. and
                   {Sip{\H{o}}cz}, B.~M. and {G{\"u}nther}, H.~M. and
                   {Lim}, P.~L. and {Crawford}, S.~M. and {Conseil}, S. and
                   {Shupe}, D.~L. and {Craig}, M.~W. and {Dencheva}, N. and
                   {Ginsburg}, A. and {VanderPlas}, J.~T. and
                   {Bradley}, L.~D. and {P{\'e}rez-Su{\'a}rez}, D. and
                   {de Val-Borro}, M. and {Aldcroft}, T.~L. and
                   {Cruz}, K.~L. and {Robitaille}, T.~P. and
                   {Tollerud}, E.~J. and {Ardelean}, C. and {Babej}, T. and
                   {Bach}, Y.~P. and {Bachetti}, M. and {Bakanov}, A.~V. and
                   {Bamford}, S.~P. and {Barentsen}, G. and {Barmby}, P. and
                   {Baumbach}, A. and {Berry}, K.~L. and {Biscani}, F. and
                   {Boquien}, M. and {Bostroem}, K.~A. and
                   {Bouma}, L.~G. and {Brammer}, G.~B. and {Bray}, E.~M. and
                   {Breytenbach}, H. and {Buddelmeijer}, H. and
                   {Burke}, D.~J. and {Calderone}, G. and
                   {Cano Rodr{\'\i}guez}, J.~L. and {Cara}, M. and
                   {Cardoso}, J.~V.~M. and {Cheedella}, S. and
                   {Copin}, Y. and {Corrales}, L. and {Crichton}, D. and
                   {D'Avella}, D. and {Deil}, C. and {Depagne}, {\'E}. and
                   {Dietrich}, J.~P. and {Donath}, A. and
                   {Droettboom}, M. and {Earl}, N. and {Erben}, T. and
                   {Fabbro}, S. and {Ferreira}, L.~A. and {Finethy}, T. and
                   {Fox}, R.~T. and {Garrison}, L.~H. and
                   {Gibbons}, S.~L.~J. and {Goldstein}, D.~A. and
                   {Gommers}, R. and {Greco}, J.~P. and {Greenfield}, P. and
                   {Groener}, A.~M. and {Grollier}, F. and {Hagen}, A. and
                   {Hirst}, P. and {Homeier}, D. and {Horton}, A.~J. and
                   {Hosseinzadeh}, G. and {Hu}, L. and {Hunkeler}, J.~S. and
                   {Ivezi{\'c}}, {\v{Z}}. and {Jain}, A. and
                   {Jenness}, T. and {Kanarek}, G. and {Kendrew}, S. and
                   {Kern}, N.~S. and {Kerzendorf}, W.~E. and
                   {Khvalko}, A. and {King}, J. and {Kirkby}, D. and
                   {Kulkarni}, A.~M. and {Kumar}, A. and {Lee}, A. and
                   {Lenz}, D. and {Littlefair}, S.~P. and {Ma}, Z. and
                   {Macleod}, D.~M. and {Mastropietro}, M. and
                   {McCully}, C. and {Montagnac}, S. and {Morris}, B.~M. and
                   {Mueller}, M. and {Mumford}, S.~J. and {Muna}, D. and
                   {Murphy}, N.~A. and {Nelson}, S. and {Nguyen}, G.~H. and
                   {Ninan}, J.~P. and {N{\"o}the}, M. and {Ogaz}, S. and
                   {Oh}, S. and {Parejko}, J.~K. and {Parley}, N. and
                   {Pascual}, S. and {Patil}, R. and {Patil}, A.~A. and
                   {Plunkett}, A.~L. and {Prochaska}, J.~X. and
                   {Rastogi}, T. and {Reddy Janga}, V. and {Sabater}, J. and
                   {Sakurikar}, P. and {Seifert}, M. and
                   {Sherbert}, L.~E. and {Sherwood-Taylor}, H. and
                   {Shih}, A.~Y. and {Sick}, J. and {Silbiger}, M.~T. and
                   {Singanamalla}, S. and {Singer}, L.~P. and
                   {Sladen}, P.~H. and {Sooley}, K.~A. and
                   {Sornarajah}, S. and {Streicher}, O. and {Teuben}, P. and
                   {Thomas}, S.~W. and {Tremblay}, G.~R. and
                   {Turner}, J.~E.~H. and {Terr{\'o}n}, V. and
                   {van Kerkwijk}, M.~H. and {de la Vega}, A. and
                   {Watkins}, L.~L. and {Weaver}, B.~A. and
                   {Whitmore}, J.~B. and {Woillez}, J. and {Zabalza}, V. and
                   {Astropy Contributors}},
  journal =       {\aj},
  month =         sep,
  number =        {3},
  pages =         {123},
  title =         {{The Astropy Project: Building an Open-science
                   Project and Status of the v2.0 Core Package}},
  volume =        {156},
  year =          {2018},
  doi =           {10.3847/1538-3881/aabc4f},
  eid =           {123},
}

@article{2022ApJ...935..167A,
  author =        {{Astropy Collaboration} and {Price-Whelan}, Adrian M. and
                   {Lim}, Pey Lian and {Earl}, Nicholas and
                   {Starkman}, Nathaniel and {Bradley}, Larry and
                   {Shupe}, David L. and {Patil}, Aarya A. and
                   {Corrales}, Lia and {Brasseur}, C.~E. and
                   {N{\"o}the}, Maximilian and {Donath}, Axel and
                   {Tollerud}, Erik and {Morris}, Brett M. and
                   {Ginsburg}, Adam and {Vaher}, Eero and
                   {Weaver}, Benjamin A. and {Tocknell}, James and
                   {Jamieson}, William and {van Kerkwijk}, Marten H. and
                   {Robitaille}, Thomas P. and {Merry}, Bruce and
                   {Bachetti}, Matteo and {G{\"u}nther}, H. Moritz and
                   {Aldcroft}, Thomas L. and {Alvarado-Montes}, Jaime A. and
                   {Archibald}, Anne M. and {B{\'o}di}, Attila and
                   {Bapat}, Shreyas and {Barentsen}, Geert and
                   {Baz{\'a}n}, Juanjo and {Biswas}, Manish and
                   {Boquien}, M{\'e}d{\'e}ric and {Burke}, D.~J. and
                   {Cara}, Daria and {Cara}, Mihai and {Conroy}, Kyle E. and
                   {Conseil}, Simon and {Craig}, Matthew W. and
                   {Cross}, Robert M. and {Cruz}, Kelle L. and
                   {D'Eugenio}, Francesco and {Dencheva}, Nadia and
                   {Devillepoix}, Hadrien A.~R. and
                   {Dietrich}, J{\"o}rg P. and {Eigenbrot}, Arthur Davis and
                   {Erben}, Thomas and {Ferreira}, Leonardo and
                   {Foreman-Mackey}, Daniel and {Fox}, Ryan and
                   {Freij}, Nabil and {Garg}, Suyog and {Geda}, Robel and
                   {Glattly}, Lauren and {Gondhalekar}, Yash and
                   {Gordon}, Karl D. and {Grant}, David and
                   {Greenfield}, Perry and {Groener}, Austen M. and
                   {Guest}, Steve and {Gurovich}, Sebastian and
                   {Handberg}, Rasmus and {Hart}, Akeem and
                   {Hatfield-Dodds}, Zac and {Homeier}, Derek and
                   {Hosseinzadeh}, Griffin and {Jenness}, Tim and
                   {Jones}, Craig K. and {Joseph}, Prajwel and
                   {Kalmbach}, J. Bryce and {Karamehmetoglu}, Emir and
                   {Ka{\l}uszy{\'n}ski}, Miko{\l}aj and
                   {Kelley}, Michael S.~P. and {Kern}, Nicholas and
                   {Kerzendorf}, Wolfgang E. and {Koch}, Eric W. and
                   {Kulumani}, Shankar and {Lee}, Antony and {Ly}, Chun and
                   {Ma}, Zhiyuan and {MacBride}, Conor and
                   {Maljaars}, Jakob M. and {Muna}, Demitri and
                   {Murphy}, N.~A. and {Norman}, Henrik and
                   {O'Steen}, Richard and {Oman}, Kyle A. and
                   {Pacifici}, Camilla and {Pascual}, Sergio and
                   {Pascual-Granado}, J. and {Patil}, Rohit R. and
                   {Perren}, Gabriel I. and {Pickering}, Timothy E. and
                   {Rastogi}, Tanuj and {Roulston}, Benjamin R. and
                   {Ryan}, Daniel F. and {Rykoff}, Eli S. and
                   {Sabater}, Jose and {Sakurikar}, Parikshit and
                   {Salgado}, Jes{\'u}s and {Sanghi}, Aniket and
                   {Saunders}, Nicholas and {Savchenko}, Volodymyr and
                   {Schwardt}, Ludwig and {Seifert-Eckert}, Michael and
                   {Shih}, Albert Y. and {Jain}, Anany Shrey and
                   {Shukla}, Gyanendra and {Sick}, Jonathan and
                   {Simpson}, Chris and {Singanamalla}, Sudheesh and
                   {Singer}, Leo P. and {Singhal}, Jaladh and
                   {Sinha}, Manodeep and {Sip{\H{o}}cz}, Brigitta M. and
                   {Spitler}, Lee R. and {Stansby}, David and
                   {Streicher}, Ole and {{\v{S}}umak}, Jani and
                   {Swinbank}, John D. and {Taranu}, Dan S. and
                   {Tewary}, Nikita and {Tremblay}, Grant R. and
                   {de Val-Borro}, Miguel and {Van Kooten}, Samuel J. and
                   {Vasovi{\'c}}, Zlatan and {Verma}, Shresth and
                   {de Miranda Cardoso}, Jos{\'e} Vin{\'\i}cius and
                   {Williams}, Peter K.~G. and {Wilson}, Tom J. and
                   {Winkel}, Benjamin and {Wood-Vasey}, W.~M. and
                   {Xue}, Rui and {Yoachim}, Peter and {Zhang}, Chen and
                   {Zonca}, Andrea and {Astropy Project Contributors}},
  journal =       {\apj},
  month =         aug,
  number =        {2},
  pages =         {167},
  title =         {{The Astropy Project: Sustaining and Growing a
                   Community-oriented Open-source Project and the Latest
                   Major Release (v5.0) of the Core Package}},
  volume =        {935},
  year =          {2022},
  doi =           {10.3847/1538-4357/ac7c74},
  eid =           {167},
}

@article{2015arXiv150708343V,
  author =        {{Vianello}, Giacomo and {Lauer}, Robert J. and
                   {Younk}, Patrick and {Tibaldo}, Luigi and
                   {Burgess}, James M. and {Ayala}, Hugo and
                   {Harding}, Patrick and {Hui}, Michelle and
                   {Omodei}, Nicola and {Zhou}, Hao},
  journal =       {arXiv e-prints},
  month =         jul,
  pages =         {arXiv:1507.08343},
  title =         {{The Multi-Mission Maximum Likelihood framework
                   (3ML)}},
  year =          {2015},
  doi =           {10.48550/arXiv.1507.08343},
  eid =           {arXiv:1507.08343},
}

@article{Gill2026,
	title = {Prospects of prompt gamma-ray burst polarimetry with {POLAR}-2},
	volume = {547},
	issn = {0035-8711},
	url = {https://doi.org/10.1093/mnras/stag452},
	doi = {10.1093/mnras/stag452},
	number = {3},
	urldate = {2026-05-10},
	journal = {Monthly Notices of the Royal Astronomical Society},
	author = {Gill, Ramandeep and He, Jiang and Granot, Jonathan and Sun, Jian-Chao and Zhang, Shuang-Nan and Wang, Yuan-Hao and Hulsman, Johannes and Produit, Nicolas and Xiong, Shao-Lin},
	month = apr,
	year = {2026},
	pages = {stag452},
}

@Article{numpy2020,
 title         = {Array programming with {NumPy}},
 author        = {Charles R. Harris and K. Jarrod Millman and St{\'{e}}fan J.
                 van der Walt and Ralf Gommers and Pauli Virtanen and David
                 Cournapeau and Eric Wieser and Julian Taylor and Sebastian
                 Berg and Nathaniel J. Smith and Robert Kern and Matti Picus
                 and Stephan Hoyer and Marten H. van Kerkwijk and Matthew
                 Brett and Allan Haldane and Jaime Fern{\'{a}}ndez del
                 R{\'{i}}o and Mark Wiebe and Pearu Peterson and Pierre
                 G{\'{e}}rard-Marchant and Kevin Sheppard and Tyler Reddy and
                 Warren Weckesser and Hameer Abbasi and Christoph Gohlke and
                 Travis E. Oliphant},
 year          = {2020},
 month         = sep,
 journal       = {Nature},
 volume        = {585},
 number        = {7825},
 pages         = {357--362},
 doi           = {10.1038/s41586-020-2649-2},
 publisher     = {Springer Science and Business Media {LLC}},
 url           = {https://doi.org/10.1038/s41586-020-2649-2}
}

@Article{matplotlib2007,
  Author    = {Hunter, J. D.},
  Title     = {Matplotlib: A 2D graphics environment},
  Journal   = {Computing in Science \& Engineering},
  Volume    = {9},
  Number    = {3},
  Pages     = {90--95},
  publisher = {IEEE COMPUTER SOC},
  doi       = {10.1109/MCSE.2007.55},
  year      = 2007
}

@ARTICLE{scipy2020,
  author  = {Virtanen, Pauli and Gommers, Ralf and Oliphant, Travis E. and
            Haberland, Matt and Reddy, Tyler and Cournapeau, David and
            Burovski, Evgeni and Peterson, Pearu and Weckesser, Warren and
            Bright, Jonathan and {van der Walt}, St{\'e}fan J. and
            Brett, Matthew and Wilson, Joshua and Millman, K. Jarrod and
            Mayorov, Nikolay and Nelson, Andrew R. J. and Jones, Eric and
            Kern, Robert and Larson, Eric and Carey, C J and
            Polat, {\.I}lhan and Feng, Yu and Moore, Eric W. and
            {VanderPlas}, Jake and Laxalde, Denis and Perktold, Josef and
            Cimrman, Robert and Henriksen, Ian and Quintero, E. A. and
            Harris, Charles R. and Archibald, Anne M. and
            Ribeiro, Ant{\^o}nio H. and Pedregosa, Fabian and
            {van Mulbregt}, Paul and {SciPy 1.0 Contributors}},
  title   = {{{SciPy} 1.0: Fundamental Algorithms for Scientific
            Computing in Python}},
  journal = {Nature Methods},
  year    = {2020},
  volume  = {17},
  pages   = {261--272},
  adsurl  = {https://rdcu.be/b08Wh},
  doi     = {10.1038/s41592-019-0686-2},
}

@ARTICLE{fermi10yr,
       author = {{Poolakkil}, S. and {Preece}, R. and {Fletcher}, C. and {Goldstein}, A. and {Bhat}, P.~N. and {Bissaldi}, E. and {Briggs}, M.~S. and {Burns}, E. and {Cleveland}, W.~H. and {Giles}, M.~M. and {Hui}, C.~M. and {Kocevski}, D. and {Lesage}, S. and {Mailyan}, B. and {Malacaria}, C. and {Paciesas}, W.~S. and {Roberts}, O.~J. and {Veres}, P. and {von Kienlin}, A. and {Wilson-Hodge}, C.~A.},
        title = "{The Fermi-GBM Gamma-Ray Burst Spectral Catalog: 10 yr of Data}",
      journal = {\apj},
         year = 2021,
        month = may,
       volume = {913},
       number = {1},
          eid = {60},
        pages = {60},
          doi = {10.3847/1538-4357/abf24d},
archivePrefix = {arXiv},
       eprint = {2103.13528},
 primaryClass = {astro-ph.HE},
       adsurl = {https://ui.adsabs.harvard.edu/abs/2021ApJ...913...60P}
}

@ARTICLE{2019NatAs...3..258Z,
       author = {{Zhang}, Shuang-Nan and {Kole}, Merlin and {Bao}, Tian-Wei and {Batsch}, Tadeusz and {Bernasconi}, Tancredi and {Cadoux}, Franck and {Chai}, Jun-Ying and {Dai}, Zi-Gao and {Dong}, Yong-Wei and {Gauvin}, Neal and {Hajdas}, Wojtek and {Lan}, Mi-Xiang and {Li}, Han-Cheng and {Li}, Lu and {Li}, Zheng-Heng and {Liu}, Jiang-Tao and {Liu}, Xin and {Marcinkowski}, Radoslaw and {Produit}, Nicolas and {Orsi}, Silvio and {Pohl}, Martin and {Rybka}, Dominik and {Shi}, Hao-Li and {Song}, Li-Ming and {Sun}, Jian-Chao and {Szabelski}, Jacek and {Tymieniecka}, Teresa and {Wang}, Rui-Jie and {Wang}, Yuan-Hao and {Wen}, Xing and {Wu}, Bo-Bing and {Wu}, Xin and {Wu}, Xue-Feng and {Xiao}, Hua-Lin and {Xiong}, Shao-Lin and {Zhang}, Lai-Yu and {Zhang}, Li and {Zhang}, Xiao-Feng and {Zhang}, Yong-Jie and {Zwolinska}, Anna},
        title = "{Detailed polarization measurements of the prompt emission of five gamma-ray bursts}",
      journal = {Nature Astronomy},
         year = 2019,
        month = jan,
       volume = {3},
        pages = {258-264},
          doi = {10.1038/s41550-018-0664-0},
archivePrefix = {arXiv},
       eprint = {1901.04207},
 primaryClass = {astro-ph.HE},
       adsurl = {https://ui.adsabs.harvard.edu/abs/2019NatAs...3..258Z}
}
\bibliographystyle{aasjournal}



\end{document}